\documentclass[twocolumn,aps,superscriptaddress,longbibliography,pra]{revtex4-1}

\usepackage{bm}%
\usepackage{float}
\expandafter\ifx\csname package@font\endcsname\relax\else
 \expandafter\expandafter
 \expandafter\usepackage
 \expandafter\expandafter
 \expandafter{\csname package@font\endcsname}%
\fi

\usepackage{array}
\usepackage{amsmath,amssymb}
\usepackage{MnSymbol}
\usepackage{tikz-feynman,contour} 
\usetikzlibrary{shapes,arrows,positioning,automata,backgrounds,calc,er,patterns}
\usepackage{feynmf}
\tikzfeynmanset{compat=1.0.0} 
 
\usepackage{graphicx}
\usepackage{mathrsfs}
\usepackage{bm}
\usepackage{mathtools}
\usepackage{epstopdf}
\usepackage{hyperref}

\hypersetup{%
   pdfpagemode=None, 
   pdfstartpage=1,
   pdfmenubar=true,
   pdftoolbar=true,
   colorlinks = true,
   linkcolor=blue,
   citecolor=blue,
   urlcolor=blue,
   bookmarksopen=false
 }
 \usepackage{amsmath}
\usepackage{amsfonts}
\usepackage{mathtools}
\usepackage{graphicx}
\usepackage{fixme}
\usepackage{color}
\usepackage{cancel}
\usepackage{bm}

\usepackage{epstopdf}
\usepackage{float}
\usepackage{amsmath}
\usepackage{color}
\usepackage{comment}
\usepackage{dsfont}
\usepackage{braket}
\usepackage[justification=justified, format=plain]{subcaption}
\graphicspath{{./images/}}
\usepackage[justification=justified ,singlelinecheck=off]{caption}
\usepackage[normalem]{ulem}

\usepackage{feynmf}
\usepackage{tikz-feynman}
\usetikzlibrary{shapes,arrows,positioning,automata,backgrounds,calc,er,patterns}
\tikzfeynmanset{compat=1.0.0}

\definecolor{airforceblue}{rgb}{0.36, 0.54, 0.66}
\definecolor{amber}{rgb}{1.0, 0.75, 0.0}
\definecolor{applegreen}{rgb}{0.55, 0.71, 0.0}
\definecolor{alizarin}{rgb}{0.82, 0.1, 0.26}

\newcommand{\kristian}[1]{{\color{blue}#1}}

\newcommand{\Up}{{\uparrow}}
\newcommand{\Dn}{{\downarrow}}

\newcommand\bwt{\begin{widetext}}
\newcommand\ewt{\end{widetext}}

\def\rd{{\rm d}}
\def\bk{{\bf k}}
\def\bp{{\bf p}}

\def\bi{{\bf i}}
\def\bj{{\bf j}}
\def\bQ{{\bf Q}}
\def\rG{G}

\def\br{{\bf r}}
\def\ro{{\rm o}}
\def\rp{{\rm p}}
\def\bdelta{{\pmb \delta}}

\begin{document}





\title{Equilibrium and non-equilibrium dynamics of a hole in a bilayer antiferromagnet}
\author{Jens H.\ Nyhegn}
\affiliation{Center for Complex Quantum Systems, Department of Physics and Astronomy, Aarhus University, Ny Munkegade, DK-8000 Aarhus C, Denmark. }
\author{Kristian K.\ Nielsen}
\affiliation{Max-Planck Institute for Quantum Optics, Hans-Kopfermann-Str. 1, D-85748 Garching, Germany}
\affiliation{Center for Complex Quantum Systems, Department of Physics and Astronomy, Aarhus University, Ny Munkegade, DK-8000 Aarhus C, Denmark. }
\author{Georg M.\ Bruun}
\email[]{bruungmb@phys.au.dk}
\affiliation{Center for Complex Quantum Systems, Department of Physics and Astronomy, Aarhus University, Ny Munkegade, DK-8000 Aarhus C, Denmark. }
\affiliation{Shenzhen Institute for Quantum Science and Engineering and Department of Physics, Southern University of Science and Technology, Shenzhen 518055, China}

\begin{abstract}
The dynamics of charge carriers  in lattices of quantum spins  is a long standing and fundamental problem. Recently, a new generation of quantum simulation experiments based on atoms in optical lattices has emerged that gives unprecedented  insights into the detailed spatial and temporal dynamics of this problem, which compliments earlier results  from condensed matter experiments. Focusing on observables accessible in these new experiments, we explore here the equilibrium as well as non-equilibrium dynamics of a mobile
hole in  two coupled  antiferromagnetic spin lattices. Using a self-consistent Born approximation, we calculate the spectral properties of the hole in the bilayer and extract the energy bands of the quasiparticles, corresponding to magnetic polarons that 
are either symmetric or anti-symmetric under layer exchange. These two kinds of polarons are degenerate at certain momenta due to the antiferromagnetic symmetry, and we, furthermore, examine how
the momentum of the ground state polaron depends on the interlayer coupling strength.  
The long time  dynamics of a hole initially created in one layer is shown to be characterised by oscillations between the two layers with a 
  frequency given by the energy difference between the symmetric and the anti-symmetric polaron. 
  We finally demonstrate   that the expansion velocity of a hole initially created at a given lattice 
  site is governed by the ballistic   motion of polarons.  It moreover depends non-monotonically on the interlayer coupling, 
  eventually increasing as a quantum phase transition to a disordered state is approached. 
   \end{abstract}

\date{\today}
\maketitle

\section{Introduction}
The motion of  charge carriers in doped antiferromagnetic (AF) layers is a key problem in quantum many-body physics that has been studied intensely since the discovery of high temperature superconductivity. In the cuprates~\cite{Keimer:2015wh} as well as in other strongly correlated two-dimensional (2D) materials such as the pnictides~\cite{Wen2011}, organic layers~\cite{Wosnitza2012}, and twisted bilayer graphene~\cite{Cao:2018wy}, pairing exists close to the magnetically ordered phase. The competition between hole motion and AF order, therefore, provides important clues for the physics of these unconventional  superconductors~\cite{Dagotto1994,Lee2006}. 
For small hole doping, the hole dynamics  is described by the ubiquitous $t$-$J$ model, which  is able to quantitatively explain photoemission 
experiments from insulating cuprates when treated within the so-called self-consistent Born approximation (SCBA)~\cite{Damascelli2003}. 
The dressing of the hole by magnetic frustration in its vicinity leads to the formation of quasiparticles coined magnetic 
polarons~\cite{schmitt-rink1988,shraiman1988,kane1989,martinez1991,liu1991}, which play a 
key role for understanding the properties of doped AF layers. It is well-known that the SCBA yields an accurate description of  the equilibrium 
properties of the magnetic polaron~\cite{martinez1991,liu1991,marsiglio1991,chernyshev1999,Diamantis_2021}, and recently this  
was shown to hold for the non-equilibrium properties as well~\cite{nielsen2022}. 
Since the unit cell of some cuprates can have two or more  CuO$_2$ planes, magnetic polarons have also been studied in multi-layer systems~\cite{nazarenko1996,yin1997,yin1998}.
Such systems are also  interesting because they exhibit a quantum phase transition between long range AF order and a disordered 
state of spin singlets across the layers~\cite{Hida1992,Sandvik1994,scalettar1994,millis1994,Sandvik1995,chubukov1995}. 

Mobile charge carriers in two-dimensional (2D) quantum AF magnets has recently received renewed interest  due to a novel generation of  
experiments using ultracold  atoms in optical lattices~\cite{Carlstrom2016,Nagy2017,grusdt2018,Grusdt2018b,nielsen2021,nielsen2022}. 
These experiments  provide an essentially perfect realization of the Fermi-Hubbard model and combined with their single site resolution imaging, they represent a powerful quantum simulator for exploring the interplay between charge carriers and magnetic order in doped 
AFs~\cite{Christie2019,Brown2019,Koepsell:2019ua,Ji2021,Koepsell2021}. Recently,  also 
  bilayer~\cite{gall2021a} and ladder geometries~\cite{Hirthe2022} have been realised with optical lattices. 
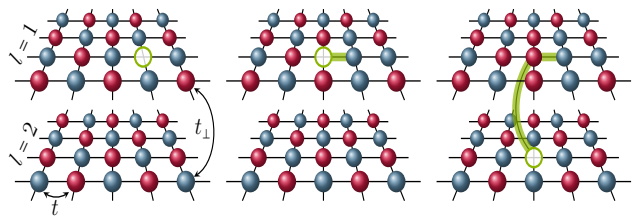
\begin{figure}[t!]
	\centering
	\hspace{-0.4cm}
	\begin{subfigure}[b]{0.15\textwidth}
		\centering
		\resizebox{1.1\textwidth}{1.05\textwidth}{%
		\begin{tikzpicture}
			\newcommand \x {2.1};
			\newcommand \dis {0.45};
			\newcommand \diss {0.4};
			\newcommand \disss {0.35};
			\newcommand \dissss {0.2};

			\draw[thick] (0,0.1) -- (0,2) ;
			\draw[thick] (-1,0.1) -- (-0.5,2) ;
			\draw[thick] (1,0.1) -- (0.5,2) ;
			\draw[thick] (-2,0.1) -- (-1.1,2) ;
			\draw[thick] (2,0.1) -- (1.1,2) ;
			\draw[thick] (0,0.1-\x) -- (0,2-\x) ;
			\draw[thick] (-1,0.1-\x) -- (-0.5,2-\x) ;
			\draw[thick] (1,0.1-\x) -- (0.5,2-\x) ;
			\draw[thick] (-2,0.1-\x) -- (-1.1,2-\x) ;
			\draw[thick] (2,0.1-\x) -- (1.1,2-\x) ;
		
		
			\draw[thick] (-2.4,+0.5-\x) -- (2.4,+0.5-\x) ;
			\draw[thick] (-2.1,+1-\x) -- (2.1,+1-\x) ;
			\draw[thick] (-1.8,+1.4-\x) -- (1.8,+1.4-\x) ;
			\draw[thick] (-1.6,+1.75-\x) -- (1.6,+1.75-\x) ;
				
			\draw[thick] (-2.4,+0.5) -- (2.4,+0.5) ;
			\draw[thick] (-2.1,+1) -- (2.1,+1) ;
			\draw[thick] (-1.8,+1.4) -- (1.8,+1.4) ;
			\draw[thick] (-1.6,+1.75) -- (1.6,+1.75) ;
			
			\filldraw[color=applegreen, fill=applegreen!0, fill opacity=0.8, line width=0.5mm] (0.75,1) circle (0.2) ;
			
			\node[circle, shading=ball, ball color=alizarin, minimum width= \dis  cm] (ball) at (0,0.5) {}; 
			\node[circle, shading=ball, ball color=airforceblue, minimum width=\diss cm] (ball) at (0,1) {}; 	
			\node[circle, shading=ball, ball color=alizarin, minimum width=\disss cm] (ball) at (0,1.4) {};
			\node[circle, shading=ball, ball color=airforceblue, minimum width=\dissss cm] (ball) at (0,1.77) {};

			\node[circle, shading=ball, ball color=airforceblue, minimum width= \dis cm] (ball) at (0.9,0.5) {};
			\node[circle, shading=ball, ball color=airforceblue, minimum width=\disss cm] (ball) at (0.65,1.4) {};
			\node[circle, shading=ball, ball color=alizarin, minimum width=\dissss cm] (ball) at (0.57,1.77) {};
		
			\node[circle, shading=ball, ball color=airforceblue, minimum width=\dis cm] (ball) at (-0.9,0.5) {};
			\node[circle, shading=ball, ball color=alizarin, minimum width=\diss cm] (ball) at (-0.75,1) {};
			\node[circle, shading=ball, ball color=airforceblue, minimum width=\disss  cm] (ball) at (-0.65,1.4) {};
			\node[circle, shading=ball, ball color=alizarin, minimum width=\dissss cm] (ball) at (-0.57,1.77) {};
		
			\node[circle, shading=ball, ball color=alizarin, minimum width=\dis cm] (ball) at (1.8,0.5) {};
			\node[circle, shading=ball, ball color=airforceblue, minimum width=\diss cm] (ball) at (1.6,1) {};
			\node[circle, shading=ball, ball color=alizarin, minimum width=\disss cm] (ball) at (1.4,1.4) {};
			\node[circle, shading=ball, ball color=airforceblue, minimum width=\dissss cm] (ball) at (1.23,1.77) {};

			\node[circle, shading=ball, ball color=alizarin, minimum width=\dis cm] (ball) at (-1.8,0.5) {};
			\node[circle, shading=ball, ball color=airforceblue, minimum width=\diss cm] (ball) at (-1.6,1) {};
			\node[circle, shading=ball, ball color=alizarin, minimum width=\disss cm] (ball) at (-1.4,1.4) {};
			\node[circle, shading=ball, ball color=airforceblue, minimum width=\dissss cm] (ball) at (-1.23,1.77) {};
		
			\node[circle, shading=ball, ball color=airforceblue, minimum width=\dis cm] (ball) at (0,0.5-\x) {};
			\node[circle, shading=ball, ball color=alizarin, minimum width=\diss cm] (ball) at (0,1-\x) {};
			\node[circle, shading=ball, ball color=airforceblue, minimum width=\disss cm] (ball) at (0,1.4-\x) {};
			\node[circle, shading=ball, ball color=alizarin, minimum width=\dissss cm] (ball) at (0,1.77-\x) {};
			
			\node[circle, shading=ball, ball color=alizarin, minimum width=\dis cm] (ball) at (0.9,0.5-\x) {};
			\node[circle, shading=ball, ball color=airforceblue, minimum width=\diss cm] (ball) at (0.75,1-\x) {};
			\node[circle, shading=ball, ball color=alizarin, minimum width=\disss cm] (ball) at (0.65,1.4-\x) {};
			\node[circle, shading=ball, ball color=airforceblue, minimum width=\dissss cm] (ball) at (0.57,1.77-\x) {};
		
			\node[circle, shading=ball, ball color=alizarin, minimum width=\dis cm] (ball) at (-0.9,0.5-\x) {};
			\node[circle, shading=ball, ball color=airforceblue, minimum width=\diss cm] (ball) at (-0.75,1-\x) {};
			\node[circle, shading=ball, ball color=alizarin, minimum width=\disss cm] (ball) at (-0.65,1.4-\x) {};
			\node[circle, shading=ball, ball color=airforceblue, minimum width=\dissss cm] (ball) at (-0.57,1.77-\x) {};	
		
			\node[circle, shading=ball, ball color=airforceblue, minimum width=\dis cm] (ball) at (1.8,0.5-\x) {};
			\node[circle, shading=ball, ball color=alizarin, minimum width=\diss cm] (ball) at (1.6,1-\x) {};
			\node[circle, shading=ball, ball color=airforceblue, minimum width=\disss cm] (ball) at (1.4,1.4-\x) {};
			\node[circle, shading=ball, ball color=alizarin, minimum width=\dissss cm] (ball) at (1.23,1.77-\x) {};

			\node[circle, shading=ball, ball color=airforceblue, minimum width=\dis cm] (ball) at (-1.8,0.5-\x) {};
			\node[circle, shading=ball, ball color=alizarin, minimum width=\diss cm] (ball) at (-1.6,1-\x) {};
			\node[circle, shading=ball, ball color=airforceblue, minimum width=\disss cm] (ball) at (-1.4,1.4-\x) {};
			\node[circle, shading=ball, ball color=alizarin, minimum width=\dissss cm] (ball) at (-1.23,1.77-\x) {};
			
			\draw [<->,thick,>=stealth] (+2,0.35) to [out=-30,in=30] (+2,0.6-\x);
			\draw [<->,thick,>=stealth] (-1.7,0.28-\x)  to [out=-30,in=210] (-1.07,0.28-\x);
		
			\node[text width=0.1cm] at (-1.45,-0.05-\x)	{\Large$t$};
			\node[text width=0.1cm] at (2.07,0.55-\x/2)	{\Large$t_{\perp}$};
		
			\node[minimum width=0.2cm, rotate=62] at (-2.2,1.3)	{\Large $l = 1$};
			\node[minimum width=0.2cm, rotate=62] at (-2.2,1.3-\x){ \Large $l = 2$};		
		\end{tikzpicture}
		}
	\end{subfigure}
	\begin{subfigure}[b]{0.15\textwidth}
		\centering
		\resizebox{1.1\textwidth}{1.05\textwidth}{%
		\begin{tikzpicture}	
			\newcommand \x {2.1};
			\newcommand \dis {0.45};
			\newcommand \diss {0.4};
			\newcommand \disss {0.35};
			\newcommand \dissss {0.2};

			\draw[thick] (0,0.1) -- (0,2) ;
			\draw[thick] (-1,0.1) -- (-0.5,2) ;
			\draw[thick] (1,0.1) -- (0.5,2) ;
			\draw[thick] (-2,0.1) -- (-1.1,2) ;
			\draw[thick] (2,0.1) -- (1.1,2) ;
			\draw[thick] (0,0.1-\x) -- (0,2-\x) ;
			\draw[thick] (-1,0.1-\x) -- (-0.5,2-\x) ;
			\draw[thick] (1,0.1-\x) -- (0.5,2-\x) ;
			\draw[thick] (-2,0.1-\x) -- (-1.1,2-\x) ;
			\draw[thick] (2,0.1-\x) -- (1.1,2-\x) ;
		
			\draw[thick] (-2.4,+0.5-\x) -- (2.4,+0.5-\x) ;
			\draw[thick] (-2.1,+1-\x) -- (2.1,+1-\x) ;
			\draw[thick] (-1.8,+1.4-\x) -- (1.8,+1.4-\x) ;
			\draw[thick] (-1.6,+1.75-\x) -- (1.6,+1.75-\x) ;
		
			\draw[thick] (-2.4,+0.5) -- (2.4,+0.5) ;
			\draw[thick] (-2.1,+1) -- (2.1,+1) ;
			\draw[thick] (-1.8,+1.4) -- (1.8,+1.4) ;
			\draw[thick] (-1.6,+1.75) -- (1.6,+1.75) ;
			
			\draw[line width=0.18cm, color=applegreen, opacity=0.7] (0.87,1) -- (0.2,1) ;
		
			\filldraw[color=applegreen, fill=applegreen!0, fill opacity=0.8, line width=0.5mm] (0,1) circle (0.2) ;
		
			\node[circle, shading=ball, ball color=alizarin, minimum width= \dis  cm] (ball) at (0,0.5) {}; 
			\node[circle, shading=ball, ball color=alizarin, minimum width=\disss cm] (ball) at (0,1.4) {};
			\node[circle, shading=ball, ball color=airforceblue, minimum width=\dissss cm] (ball) at (0,1.77) {};

			\node[circle, shading=ball, ball color=airforceblue, minimum width= \dis cm] (ball) at (0.9,0.5) {};
			\node[circle, shading=ball, ball color=airforceblue, minimum width=\diss cm] (ball) at (0.75,1) {};
			\node[circle, shading=ball, ball color=airforceblue, minimum width=\disss cm] (ball) at (0.65,1.4) {};
			\node[circle, shading=ball, ball color=alizarin, minimum width=\dissss cm] (ball) at (0.57,1.77) {};
		
			\node[circle, shading=ball, ball color=airforceblue, minimum width=\dis cm] (ball) at (-0.9,0.5) {};
			\node[circle, shading=ball, ball color=alizarin, minimum width=\diss cm] (ball) at (-0.75,1) {};
			\node[circle, shading=ball, ball color=airforceblue, minimum width=\disss  cm] (ball) at (-0.65,1.4) {};
			\node[circle, shading=ball, ball color=alizarin, minimum width=\dissss cm] (ball) at (-0.57,1.77) {};
		
			\node[circle, shading=ball, ball color=alizarin, minimum width=\dis cm] (ball) at (1.8,0.5) {};
			\node[circle, shading=ball, ball color=airforceblue, minimum width=\diss cm] (ball) at (1.6,1) {};
			\node[circle, shading=ball, ball color=alizarin, minimum width=\disss cm] (ball) at (1.4,1.4) {};
			\node[circle, shading=ball, ball color=airforceblue, minimum width=\dissss cm] (ball) at (1.23,1.77) {};

			\node[circle, shading=ball, ball color=alizarin, minimum width=\dis cm] (ball) at (-1.8,0.5) {};
			\node[circle, shading=ball, ball color=airforceblue, minimum width=\diss cm] (ball) at (-1.6,1) {};
			\node[circle, shading=ball, ball color=alizarin, minimum width=\disss cm] (ball) at (-1.4,1.4) {};
			\node[circle, shading=ball, ball color=airforceblue, minimum width=\dissss cm] (ball) at (-1.23,1.77) {};
		
			\node[circle, shading=ball, ball color=airforceblue, minimum width=\dis cm] (ball) at (0,0.5-\x) {};
			\node[circle, shading=ball, ball color=alizarin, minimum width=\diss cm] (ball) at (0,1-\x) {};
			\node[circle, shading=ball, ball color=airforceblue, minimum width=\disss cm] (ball) at (0,1.4-\x) {};
			\node[circle, shading=ball, ball color=alizarin, minimum width=\dissss cm] (ball) at (0,1.77-\x) {};
			
			\node[circle, shading=ball, ball color=alizarin, minimum width=\dis cm] (ball) at (0.9,0.5-\x) {};
			\node[circle, shading=ball, ball color=airforceblue, minimum width=\diss cm] (ball) at (0.75,1-\x) {};
			\node[circle, shading=ball, ball color=alizarin, minimum width=\disss cm] (ball) at (0.65,1.4-\x) {};
			\node[circle, shading=ball, ball color=airforceblue, minimum width=\dissss cm] (ball) at (0.57,1.77-\x) {};
		
			\node[circle, shading=ball, ball color=alizarin, minimum width=\dis cm] (ball) at (-0.9,0.5-\x) {};
			\node[circle, shading=ball, ball color=airforceblue, minimum width=\diss cm] (ball) at (-0.75,1-\x) {};
			\node[circle, shading=ball, ball color=alizarin, minimum width=\disss cm] (ball) at (-0.65,1.4-\x) {};
			\node[circle, shading=ball, ball color=airforceblue, minimum width=\dissss cm] (ball) at (-0.57,1.77-\x) {};	
		
			\node[circle, shading=ball, ball color=airforceblue, minimum width=\dis cm] (ball) at (1.8,0.5-\x) {};
			\node[circle, shading=ball, ball color=alizarin, minimum width=\diss cm] (ball) at (1.6,1-\x) {};
			\node[circle, shading=ball, ball color=airforceblue, minimum width=\disss cm] (ball) at (1.4,1.4-\x) {};
			\node[circle, shading=ball, ball color=alizarin, minimum width=\dissss cm] (ball) at (1.23,1.77-\x) {};

			\node[circle, shading=ball, ball color=airforceblue, minimum width=\dis cm] (ball) at (-1.8,0.5-\x) {};
			\node[circle, shading=ball, ball color=alizarin, minimum width=\diss cm] (ball) at (-1.6,1-\x) {};
			\node[circle, shading=ball, ball color=airforceblue, minimum width=\disss cm] (ball) at (-1.4,1.4-\x) {};
			\node[circle, shading=ball, ball color=alizarin, minimum width=\dissss cm] (ball) at (-1.23,1.77-\x) {};		
			
			\draw [<->,thick,>=stealth,white] (+2,0.35) to [out=-30,in=30] (+2,0.6-\x);
			\draw [<->,thick,>=stealth,white] (-1.71,0.3-\x)  to [out=-30,in=210] (-1.05,0.3-\x);
		
			\node[text width=0.1cm] at (-1.38,-0.05-\x)	{\Large \textcolor{white}{$t$}};
			\node[text width=0.1cm] at (2.15,0.5-\x/2)	{\Large \textcolor{white}{$t_{\perp}$}};
		
			\node[minimum width=0.2cm, rotate=58] at (-2.2,1.3)	{\Large \textcolor{white}{$l = 1$}};
			\node[minimum width=0.2cm, rotate=58] at (-2.2,1.3-\x){ \Large \textcolor{white}{$l = 2$}};		
		\end{tikzpicture}
		}
	\end{subfigure}
	\begin{subfigure}[b]{0.15\textwidth}
		\centering
		\resizebox{1.1\textwidth}{1.05\textwidth}{%
		\begin{tikzpicture}
			\newcommand \x {2.1};
			\newcommand \dis {0.45};
			\newcommand \diss {0.4};
			\newcommand \disss {0.35};
			\newcommand \dissss {0.2};

			\draw[thick] (0,0.1) -- (0,2) ;
			\draw[thick] (-1,0.1) -- (-0.5,2) ;
			\draw[thick] (1,0.1) -- (0.5,2) ;
			\draw[thick] (-2,0.1) -- (-1.1,2) ;
			\draw[thick] (2,0.1) -- (1.1,2) ;
			\draw[thick] (0,0.1-\x) -- (0,2-\x) ;
			\draw[thick] (-1,0.1-\x) -- (-0.5,2-\x) ;
			\draw[thick] (1,0.1-\x) -- (0.5,2-\x) ;
			\draw[thick] (-2,0.1-\x) -- (-1.1,2-\x) ;
			\draw[thick] (2,0.1-\x) -- (1.1,2-\x) ;
		
		
			\draw[thick] (-2.4,+0.5-\x) -- (2.4,+0.5-\x) ;
			\draw[thick] (-2.1,+1-\x) -- (2.1,+1-\x) ;
			\draw[thick] (-1.8,+1.4-\x) -- (1.8,+1.4-\x) ;
			\draw[thick] (-1.6,+1.75-\x) -- (1.6,+1.75-\x) ;
		
			\draw[thick] (-2.1,+1) -- (2.1,+1) ;
			\draw[thick] (-1.8,+1.4) -- (1.8,+1.4) ;
			\draw[thick] (-1.6,+1.75) -- (1.6,+1.75) ;
			
			\node[circle, shading=none, ball color=airforceblue, minimum width=\dissss cm] (ball) at (-0.57,1.77-\x) {};
			\node[circle, shading=none, ball color=airforceblue, minimum width=\disss cm] (ball) at (0,1.4-\x) {};
			
			\draw[thick]  (0,1.05)  to [out=-130,in=133]  (-0.1,1.15-\x);
			\draw[line width=0.18cm, color=applegreen, opacity=0.7]  (0,1.05)  to [out=-130,in=133] (-0.1,1.15-\x);
			\draw[line width=0.18cm, color=applegreen, opacity=0.7] (0.87,1) -- (0,1) ;
			
			\filldraw[color=applegreen, fill=applegreen!0, fill opacity=0.8, line width=0.5mm] (0,1-\x) circle (0.2) ;
			
			\draw[thick] (-2.4,+0.5) -- (2.4,+0.5) ;
			\node[circle, shading=none, ball color=alizarin, minimum width= \dis  cm] (ball) at (0,0.5) {};	
			
			\node[circle, shading=none, ball color=alizarin, minimum width=\diss cm] (ball) at (0,1) {}; 	
			\node[circle, shading=none, ball color=alizarin, minimum width=\disss cm] (ball) at (0,1.4) {};
			\node[circle, shading=none, ball color=airforceblue, minimum width=\dissss cm] (ball) at (0,1.77) {};

			\node[circle, shading=none, ball color=airforceblue, minimum width= \dis cm] (ball) at (0.9,0.5) {};
			\node[circle, shading=none, ball color=airforceblue, minimum width=\diss cm] (ball) at (0.75,1) {};
			\node[circle, shading=none, ball color=airforceblue, minimum width=\disss cm] (ball) at (0.65,1.4) {};
			\node[circle, shading=none, ball color=alizarin, minimum width=\dissss cm] (ball) at (0.57,1.77) {};
		
			\node[circle, shading=ball, ball color=airforceblue, minimum width=\dis cm] (ball) at (-0.9,0.5) {};
			\node[circle, shading=ball, ball color=alizarin, minimum width=\diss cm] (ball) at (-0.75,1) {};
			\node[circle, shading=ball, ball color=airforceblue, minimum width=\disss  cm] (ball) at (-0.65,1.4) {};
			\node[circle, shading=ball, ball color=alizarin, minimum width=\dissss cm] (ball) at (-0.57,1.77) {};
		
			\node[circle, shading=ball, ball color=alizarin, minimum width=\dis cm] (ball) at (1.8,0.5) {};
			\node[circle, shading=ball, ball color=airforceblue, minimum width=\diss cm] (ball) at (1.6,1) {};
			\node[circle, shading=ball, ball color=alizarin, minimum width=\disss cm] (ball) at (1.4,1.4) {};
			\node[circle, shading=ball, ball color=airforceblue, minimum width=\dissss cm] (ball) at (1.23,1.77) {};

			\node[circle, shading=ball, ball color=alizarin, minimum width=\dis cm] (ball) at (-1.8,0.5) {};
			\node[circle, shading=ball, ball color=airforceblue, minimum width=\diss cm] (ball) at (-1.6,1) {};
			\node[circle, shading=ball, ball color=alizarin, minimum width=\disss cm] (ball) at (-1.4,1.4) {};
			\node[circle, shading=ball, ball color=airforceblue, minimum width=\dissss cm] (ball) at (-1.23,1.77) {};
		
			\node[circle, shading=ball, ball color=airforceblue, minimum width=\dis cm] (ball) at (0,0.5-\x) {};
			\node[circle, shading=ball, ball color=alizarin, minimum width=\dissss cm] (ball) at (0,1.77-\x) {};
			
			\node[circle, shading=ball, ball color=alizarin, minimum width=\dis cm] (ball) at (0.9,0.5-\x) {};
			\node[circle, shading=ball, ball color=airforceblue, minimum width=\diss cm] (ball) at (0.75,1-\x) {};
			\node[circle, shading=ball, ball color=alizarin, minimum width=\disss cm] (ball) at (0.65,1.4-\x) {};
			\node[circle, shading=ball, ball color=airforceblue, minimum width=\dissss cm] (ball) at (0.57,1.77-\x) {};
		
			\node[circle, shading=ball, ball color=alizarin, minimum width=\dis cm] (ball) at (-0.9,0.5-\x) {};
			\node[circle, shading=ball, ball color=airforceblue, minimum width=\diss cm] (ball) at (-0.75,1-\x) {};
			\node[circle, shading=ball, ball color=alizarin, minimum width=\disss cm] (ball) at (-0.65,1.4-\x) {};
		
			\node[circle, shading=ball, ball color=airforceblue, minimum width=\dis cm] (ball) at (1.8,0.5-\x) {};
			\node[circle, shading=ball, ball color=alizarin, minimum width=\diss cm] (ball) at (1.6,1-\x) {};
			\node[circle, shading=ball, ball color=airforceblue, minimum width=\disss cm] (ball) at (1.4,1.4-\x) {};
			\node[circle, shading=ball, ball color=alizarin, minimum width=\dissss cm] (ball) at (1.23,1.77-\x) {};

			\node[circle, shading=ball, ball color=airforceblue, minimum width=\dis cm] (ball) at (-1.8,0.5-\x) {};
			\node[circle, shading=ball, ball color=alizarin, minimum width=\diss cm] (ball) at (-1.6,1-\x) {};
			\node[circle, shading=ball, ball color=airforceblue, minimum width=\disss cm] (ball) at (-1.4,1.4-\x) {};
			\node[circle, shading=ball, ball color=alizarin, minimum width=\dissss cm] (ball) at (-1.23,1.77-\x) {};		
		
			\node[circle, shading=ball, ball color=alizarin, minimum width=\diss cm] (ball) at (0,1) {}; 	

			\draw [<->,thick,>=stealth,white] (+2,0.35) to [out=-30,in=30] (+2,0.6-\x);
			\draw [<->,thick,>=stealth,white] (-1.71,0.3-\x)  to [out=-30,in=210] (-1.05,0.3-\x);
		
			\node[text width=0.1cm] at (-1.38,-0.05-\x)	{\Large \textcolor{white}{$t$}};
			\node[text width=0.1cm] at (2.15,0.5-\x/2)	{\Large \textcolor{white}{$t_{\perp}$}};
		
			\node[minimum width=0.2cm, rotate=58] at (-2.2,1.3)	{\Large \textcolor{white}{$l = 1$}};
			\node[minimum width=0.2cm, rotate=58] at (-2.2,1.3-\x){ \Large \textcolor{white}{$l = 2$}};		
			
		\end{tikzpicture}
		}
	\end{subfigure}
	\caption{{\bf A hole in an AF bilayer.} The blue and red balls represent fermions with spin up and down respectively, and the green circle the hole. Reading from left to right, we see the hole propagating in the AF background, destroying the AF order around it. The delocalization of the hole and the AF order are, therefore, two competing dynamics of the system.}
	\label{Fig:Bilayer}
\end{figure}

 
Here, we explore a single mobile hole in a bilayer system consisting of two square lattices of spins with AF order. We analyze the spectral properties of the hole as a function of the system parameters  using a diagrammatic approach based on the SCBA, and we discuss the properties of the two kinds of magnetic polarons existing in the system, which are either symmetric or antisymmetric under layer exchange. Focusing  on observables that are accessible in the new optical lattice experiments, we show that these polarons give rise to intriguing non-equilibrium effects of the hole such as oscillations between the two layers, and a long time expansion velocity that first decreases and then increases with the interlayer coupling as the spins approach a quantum phase transition to a disordered state. 

\section{Model} \label{sec:model}
We consider a single mobile hole in an antiferromagnetic (AF) bilayer formed by two square lattices as illustrated in Fig.~\ref{Fig:Bilayer}. 
The dynamics of the hole is described by the  $t$-$J$ model with the Hamiltonian  
\begin{equation}
\hat{H}=\hat{H}_t+\hat{H}_J
\label{Eq:TotalH}
\end{equation}
 where 
\begin{equation}
\hat{H}_t=-t\sum_{l,\langle {\bf i}, {\bf j} \rangle,\sigma} \tilde{c}^{\dagger}_{l,{\bf i},\sigma} \tilde{c}_{l,{\bf j},\sigma}    
-t_\perp\sum_{{\bf i},\sigma} \tilde{c}^{\dagger}_{1,{\bf i},\sigma} \tilde{c}_{2,{\bf j},\sigma} +\text{h.c.} 
\label{Eq:Ht}
\end{equation}
and 
\begin{align}
\hat{H}_J = & \ J\sum_{l,\langle {\bf i}, {\bf j} \rangle } \left[ \hat{\mathbf S }_{l,{\bf i}}\cdot \hat{\mathbf S }_{l,{\bf j}} -\frac{1}{4}\hat{n}_{l,\bi}n_{l,\bj} \right] + \nonumber \\ 
& \ J_{\perp}\sum_{{\bf i}}  \left[  \hat{\mathbf S }_{1,{\bf i}}\cdot \hat{\mathbf S }_{2,{\bf i}} -\frac{1}{4}\hat{n}_{1,\bi}\hat{n}_{2,\bi} \right]
\label{Eq:HJ}
\end{align}
Here, $\tilde{c}^{\dagger}_{l,{\bf i},\sigma}=\hat {c}^{\dagger}_{l,{\bf i},\sigma}(1-\hat n_{l,{\bf i},\bar{\sigma}})$ where $\hat{c}^{\dagger}_{l,{\bf i},\sigma}$creates a fermion in layer $l=1,2$ at site $\mathbf i$ with spin $\sigma=\uparrow,\downarrow$. The factor $1-\hat n_{l,{\bf i},\bar{\sigma}}$ with $\hat n_{l,{\bf i},\bar{\sigma}}=\hat {c}^{\dagger}_{l,{\bf i},\bar{\sigma}}\hat {c}_{l,{\bf i},\bar{\sigma}}$ and $\bar{\sigma}$ the opposite spin of $\sigma$ ensures that no site is doubly occupied. The matrix elements for inter- and intralayer hopping are $t$ and $t_\perp$, and the AF coupling between neighbouring spins within the same layer and in different layers is $J>0$ and $J_\perp>0$ respectively. Also, the spin $1/2$ operators are given in terms of the fermions via the 
Schwinger  representation  
\begin{equation}
\hat{ {\bf S} }_{l,\bf i} = \frac{1}{2}\sum_{\sigma,\sigma'} \hat{ c }^\dagger_{l,{\bf i},\sigma}\boldsymbol{\sigma}_{\sigma\sigma'}\hat{ c }_{l,{\bf i},\sigma'},
\label{Eq:Schwinger_fermion}
\end{equation}
with $\boldsymbol{\sigma}=(\sigma_x,\sigma_y,\sigma_z)$ a vector of  Pauli matrices. 
Taking $J=4t^2/U$ and $J_\perp=4t_\perp^2/U$, the $t$-$J$ model provides an effective low-energy description of the Fermi-Hubbard model with strong onsite repulsion $U\gg t$ ~\cite{Chao_1977,reischl2004}, but it can also be regarded as an independent model in itself. One can for instance realize models with $J_{\perp}\neq4t_{\perp}^2/U$ using atoms in optical lattices~\cite{grusdt2018,koepsell2020} as well as using Rydberg atoms in optical tweezers~\cite{Browaeys:2020tl}.

\subsection{Slave-fermion representation}
To describe the motion of a single hole in the  bilayer, we perform a  Holstein-Primakoff transformation generalized to the case where a hole is present~\cite{kane1989,martinez1991,liu1991,schmitt-rink1988,nielsen2021}. Due to the AF order, we can for each layer define two sublattices A and B where the spins will predominantly point up in sublattice A and down in B. A site in sublattice A in layer 1 is adjacent to a site in sublattice B in layer 2 and vice versa, see Fig.~\ref{Fig:Bilayer}. For sublattice A, the spin operators are expressed as $\hat{ S }^{z}_{l,{\mathbf i}} = (1-\hat{ h }^{\dagger}_{l,{\mathbf i}}\hat{ h }_{l,{\mathbf i}})/2 - \hat{ s }^{\dagger}_{l,{\mathbf i}}\hat{ s }_{l,{\mathbf i}} $ and $\hat S_{l,{\mathbf i}}^{-}= \hat S^x_{l,{\mathbf i}}-i\hat S^y_{l,{\mathbf i}}=\hat{ s }^{\dagger}_{l,{\mathbf i}} (1-\hat{ s }^{\dagger}_{l,{\mathbf i}}\hat{ s }_{l,{\mathbf i}}-\hat{ h }^{\dagger}_{l,{\mathbf i}}\hat{ h }_{l,{\mathbf i}})^{1/2}$, and the creation operators are $\tilde{c}_{i,\Dn} = \hat{ h }^{\dagger}_{l,{\mathbf i}}\hat{ s }_{l,{\mathbf i}}$ and $\tilde{c}_{i,\Up} = \hat{ h }^{\dagger}_{l,{\mathbf i}}(1-\hat{ s }^{\dagger}_{l,{\mathbf i}}\hat{ s }_{l,{\mathbf i}}-\hat{ h }^{\dagger}_{l,{\mathbf i}}\hat{ h }_{l,{\mathbf i}})^{1/2} $. Here, $\hat{ h }^{\dagger}_{l,{\mathbf i}}$ is a fermionic creation operator of a hole and $\hat{ s }^{\dagger}_{l,{\mathbf i}}$ is a bosonic creation operator of a spin-fluctuation. The square root  $(1-\hat{ s }^{\dagger}_{l,{\mathbf i}}\hat{ s }_{l,{\mathbf i}}-\hat{ h }^{\dagger}_{l,{\mathbf i}}\hat{ h }_{l,{\mathbf i}})^{1/2}$ give a combined hardcore constraint for both spin excitations and holes. Operators on sublattice B take similar form~\cite{nielsen2021}. 

Applying this so-called  slave-fermion representation to the spin part of the Hamiltonian given by Eq.~\eqref{Eq:HJ} and keeping only linear terms  yields after diagonalization
\begin{align}
	\hat{ H  }_J = E_0+ \sum_{\bk,\mu=\pm}\omega_{\mu, \bk} \hat{ b}^{\dagger}_{\mu,\bk}\hat{ b}_{\mu,\bk}. 
\end{align}
Here, $E_0 = -{J_{\perp}}(2N-1)/{2} -{Jz}(2N-1)/{2} +\sum_{\textbf{k},\mu=\pm} \omega_{\mathbf{k},\mu}/2$ is the ground state energy 
with  $N$ the number of lattice sites in each plane, and 
\begin{align}
	\omega_{\pm,\bk} =& \ \frac{1}{2} \sqrt{\left(Jz+J_{\perp}\right)^{2}- (Jz\gamma_{\bk} \pm   J_{\perp}  )^{2}}
\label{Eq:dispersion_rel}
\end{align}
is the spin wave spectrum, with the structure factor $\gamma_{\bk} = \sum_{\bdelta} {\rm e}^{i\bk\cdot\bdelta} / z = (\cos{k_{x}} + \cos{k_{y}} )/2$. Here, $z=4$ is the coordination number and $\bdelta$ are the nearest neighbor sites, while the crystal momentum ${\bk}$ is in the first Brillouin zone (BZ) of the square lattice~\cite{nazarenko1996,yin1997,yin1998}. We take the lattice constant to be unity throughout. The bilayer has two spin wave branches $\mu=\pm$, which are connected to the spin fluctuation operators by a unitary and a Bogoliubov  
transformation 
\begin{align}
		\begin{bmatrix}
   		\hat{ s }_{1,\bk} \\ 
			\hat{ s }^{\dagger}_{1,-\bk} \\
   		\hat{ s }_{2,\bk} \\ 
			\hat{ s }^{\dagger}_{2,-\bk} 
		\end{bmatrix}
		=
		 \frac{1}{\sqrt{2}}
		\begin{bmatrix}
   			\mathds{1}  &  -\mathds{1} \\
    			\mathds{1}   &  \mathds{1}
		\end{bmatrix}
		\begin{bmatrix}
   			U_{+,\bk} & 0 \\
    			0 &  U_{-,\bk}
		\end{bmatrix}
		\begin{bmatrix}
   		\hat{ b}_{+,\bk} \\ 
			\hat{ b}^{\dagger}_{+,-\bk} \\
   		\hat{ b}_{-,\bk} \\ 
			\hat{ b}^{\dagger}_{-,-\bk} 
		\end{bmatrix}.
	\label{Eq:H_J_k}
\end{align}
Here, $\mathds{1}=\begin{bsmallmatrix}  1 & 0\\  0 & 1\end{bsmallmatrix}$ and 
$U_{\pm,\bk} = \begin{bsmallmatrix}
   			u_{\pm,\bk} & -v_{\pm,\bk} \\
    			-v_{\pm,\bk} & u_{\pm,\bk}
				\end{bsmallmatrix}
$
with the coherence factors
\begin{align}
u_{\pm,\bk} &= \sqrt{\frac{1}{2}\left(\frac{zJ + J_\perp}{2\omega_{\pm,\bk}} + 1\right)} \nonumber \\ 
v_{\pm,\bk} &= {\rm sgn}\left[ z J\gamma_\bk \pm  J_\perp\right]\sqrt{\frac{1}{2}\left(\frac{zJ + J_\perp}{2\omega_{\pm,\bk}} - 1\right)}.
\label{Eq:coherence_factors}
\end{align}
Operators in momentum space are connected to operators in real space via the usual discrete Fourier transform 
$\hat s_{l,{\bk}}=\sum_{\mathbf i}\hat s_{l,{\mathbf i}}\exp(-i{\bk}\cdot{\mathbf i})/\sqrt N$. 
The full details of the diagonalization of $\hat{ H  }_J$ can be found in Appendix \ref{app:Diagonalization}. 
Figure \ref{Fig:dispersion} shows the spin wave dispersion for the two branches taking $J_{\perp}/J = 0.25$. At low momenta, the $\mu=+$ branch 
corresponds to spin waves in the two planes being in-phase giving a Goldstone mode for $\bk\rightarrow 0$. This is reversed close to $\bp = (\pi,\pi)$
where the $\mu=-$ mode corresponds to the spin waves in the two planes being in-phase. More generally, we have $\omega_{+,\bk+\mathbf{Q}} = \omega_{-,\bk}$ where  $\mathbf{Q} = (\pi, \pi)$ is the AF ordering vector, so that the full spin wave spectrum displays the expected symmetry from the AF order.

\begin{figure}[!t]
	\centering
	\hspace{-0.4cm}
	\includegraphics[width=0.5\textwidth]{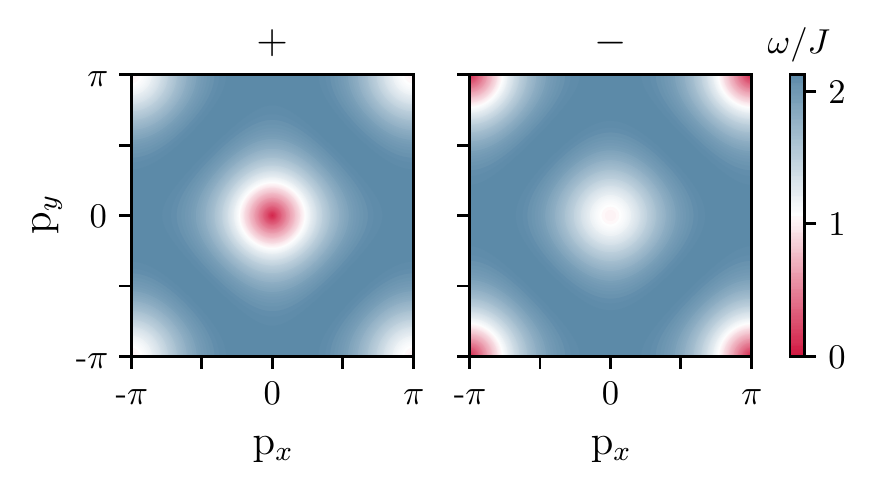}
	\caption{ {\bf Spin wave dispersion.} The left and right panel show the spin wave dispersion given by Eq. \eqref{Eq:dispersion_rel} for the $ \mu =+$ and $\mu=-$ branches respectively with $J_{\perp} /J= 0.25$.} 
	\label{Fig:dispersion}
\end{figure}

Likewise, using the slave fermion representation for $\hat{H}_t$ yields 
\begin{align}
	\hat{ H }_t= &\sum_{\bk,\bp,l}  \hat{ h }_{l,\bp+\bk}^{\dagger}\hat{ h }_{l,\bp}[g_{+}(\bp, \bk)  \hat{ b}^{\dagger}_{+,-\bk}+(-1)^{l}  g_{-}(\bp, \bk) \hat{ b}^{\dagger}_{-,-\bk}] \nonumber \\
	&+\sum_{\bk,\bp} \hat{ h }_{2,\bp+\bk}^{\dagger}\hat{ h }_{1,\bp}[ f_{+}(\bk) (\hat{ b}_{+,\bk} + \hat{ b}^{\dagger}_{+,-\bk}) \nonumber \\
	  &+ f_{-}(\bk) (\hat{ b}_{-,\bk} - \hat{ b}^{\dagger}_{-,-\bk})] + \text{h.c.}.
		\label{Eq:HtSlaveFermion}
\end{align}
Here 
\begin{align}
g_{\pm}(\bp,\bk) = \frac{zt}{\sqrt{2N}}\left[u_{\pm,\bk} \gamma_{\bp + \bk} - v_{\pm,\bk}\gamma_\bp\right]
\label{Eq:g_vertices}
\end{align}
is the vertex describing the scattering between a hole and a spin wave where the hole remains in a given layer, and 
\begin{align}
f_{\pm}(\bk) =  \frac{t_\perp}{\sqrt{2N}}\left[u_{\pm,\bk} \mp v_{\pm,\bk} \right]
\label{Eq:f_vertices}
\end{align}
is the scattering vertex when the hole jumps from one layer to the other. 
Equation \eqref{Eq:HtSlaveFermion} quantitatively describes how the motion of the hole distorts the AF order by the   
 emission of spin waves.  The inter-layer vertices $f_{\pm}(\bk)$ vanish when  $t_\perp = 0$ and the Hamiltonian then simplifies to two copies of a single layer as expected. 
 The interaction vertices satisfy the symmetries  
\begin{align}
g_{\pm}(\bp,\bk + \bQ) &= -g_{\mp}(\bp,\bk), \; g_{\pm}(\bp + \bQ,\bk) = - g_{\pm}(\bp,\bk) \nonumber \\
f_{\pm}(\bk + \bQ) &= f_{\mp}(\bk)
\label{Eq:interaction_vertex_symmetry}
\end{align}
due to the underlying AF order. 

\section{Symmetric and anti-symmetric polarons} \label{Polarons}
The competition between the hole motion and the magnetic order leads to the formation of quasiparticles where the hole is surrounded by a 
cloud of reduced AF order \cite{schmitt-rink1988,shraiman1988,kane1989,martinez1991,liu1991}. These quasiparticles, named magnetic polarons, play a 
central  role for the equilibrium as well as non-equilibrium properties of the system. 
The bilayer symmetry $\hat {c}_{1,{\bf i}}\leftrightarrow \hat {c}_{2,{\bf i}}$ of Eq.~\eqref{Eq:TotalH}, means that the polaron states can be divided into those symmetric and antisymmetric under layer exchange.
The wave function for the polarons can be expressed as a series of terms with an increasing number of spin waves on top of the AF ground state, i.e.  
\begin{align}
|\Psi_{\mathbf p}^{\pm}\rangle=& \ \sqrt{\frac{Z_{\mathbf p}^{\pm}}{2}}(\hat h^\dagger_{1,\mathbf p}\pm\hat h^\dagger_{2,\mathbf p})|\text{AF}\rangle+
\nonumber\\
& \ \sum_{{\mathbf k},\mu,l}\phi^{\pm}_{l,\mu}({\mathbf p},\mathbf{k})\hat h^\dagger_{l,{\mathbf p}+{\mathbf k}}\hat b^\dagger_{\mu,-{\mathbf k}}|\text{AF}\rangle\ldots.
\label{Eq:SymAntiSymWavefn}
\end{align}
Here,  $|\text{AF}\rangle$ is the AF ground state defined by $\hat b_{\mu,{\mathbf k}}|\text{AF}\rangle=0$, $Z_{\mathbf p}^{\pm}$ is 
the quasiparticle residue, and $\phi^{\pm}_{l,\mu}({\mathbf p},\mathbf{k})$ is the coefficient for the term involving a hole with momentum ${\mathbf p}+\mathbf{k}$ in layer $l$ and a spin wave in branch $\mu$ with momentum $-\mathbf{k}$. 
For a single layer, one has developed diagrammatic rules for constructing the wave function corresponding
to the SCBA, which was used to calculate terms including up to three spin waves \cite{reiter1994c,ramsak1998c,ramsak1993}. Recently, this has been extended to infinite order in the number of spin waves by deriving a set of Dyson like equations~\cite{nielsen2021}. 

\section{Self-consistent Born approximation} \label{SCBA}
We use the self-consistent Born approximation (SCBA) \cite{schmitt-rink1988,kane1989} generalized to the case of a bilayer to analyze the properties of the hole~\cite{nazarenko1996,yin1997,yin1998}. For a single layer, the SCBA is known to yield quantitatively accurate results for the equilibrium properties of the hole~\cite{martinez1991,liu1991,marsiglio1991,liu1992,chernyshev1999,Diamantis_2021}, and recently this has been shown to hold for the  non-equilibrium  dynamics as well~\cite{nielsen2022}.

The layer degree of freedom gives rise to a $2\times2$ matrix structure for the equilibrium Green's function for the hole, which we define as
\begin{align}
G_{lm}(\bp,\tau) = -\braket{T_{\tau} \left[ \hat{ h }_{l,\bp}(\tau)\hat{ h }^{\dagger}_{m,\bp}(0) \right]}, 
\end{align}
where $T$ is time ordering in imaginary time $\tau$. 
We have $G_{11}(\bp,\tau)=G_{22}(\bp,\tau)\equiv G_{\rm d}(\bp,\tau)$
and $G_{12}(\bp,\tau)=G_{21}(\bp,\tau)\equiv G_{\rm o}(\bp,\tau)$ due to the layer symmetry. The Dyson equation reads in frequency space 
\begin{equation}
G(p) = \mathds{1}_{2} G_0(p)  + G_0(p) \Sigma(p) G(p), 
\label{Eq:dyson_equation}
\end{equation}
where $p = (\bp, i\omega_p)$ with $i\omega_p$ a fermionic Matsubara frequency.
The non-interacting hole Green's function is $G_0(p) = 1 / i\omega_p$ independent of the crystal momentum since there is no kinetic energy term 
for the hole in $\hat{ H }_t$ given Eq.~\eqref{Eq:HtSlaveFermion}.
The self-energy matrix is
\begin{align}
\Sigma(p) = \begin{bmatrix} \Sigma_{\rm d}(p) & \Sigma_{\rm o}(p) \\ \Sigma_{\rm o}(p) & \Sigma_{\rm d}(p) \end{bmatrix}.
\label{Eq:Self_Energy_Matrix}
\end{align}
From Eq.~\eqref{Eq:dyson_equation} we find 
 \begin{align}
G_{\rm d}(p) = \frac{  i\omega_p -\Sigma_{\rm d}(p) }{ [ i\omega_p - \Sigma_{\rm d}(p)]^{2}- \Sigma_{\rm o}(p)^{2} } \nonumber \\
G_{\rm o}(p) = \frac{ \Sigma_{\rm o}(p) }{ [  i\omega_p - \Sigma_{\rm d}(p)]^{2} - \Sigma_{\rm o}(p) ^{2} }. 
	\label{Eq:Prop_D_O}
\end{align}

\begin{figure}[t!]
	\centering
	\hspace{-1cm}
	\includegraphics[width=0.8\columnwidth]{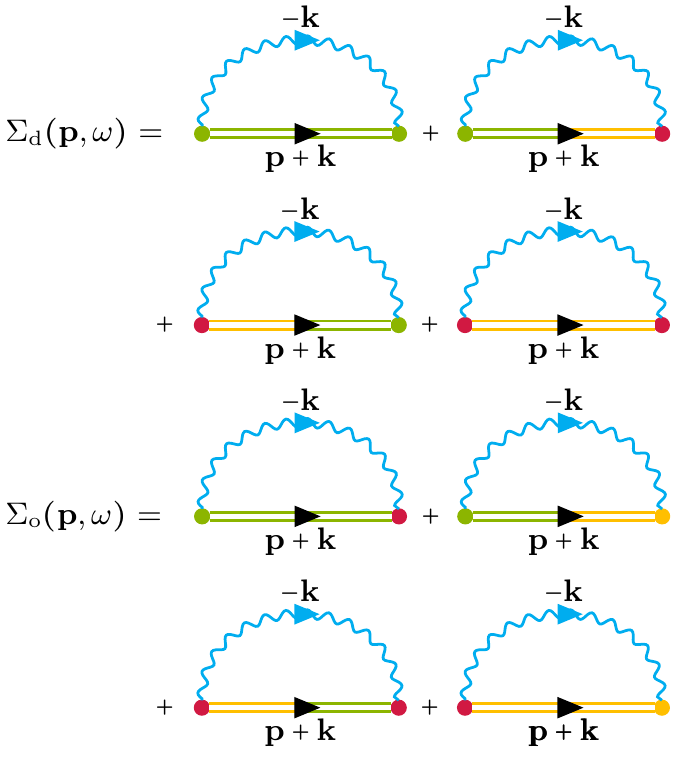}
	\caption{{\bf SCBA diagrams.} Feynman diagrams included in the calculation of $\Sigma_{\text{d}}$ and $\Sigma_{\text{o}}$. The green color is associated with layer $l=1$ and yellow with $l=2$. Green and yellow vertices are associated with the intra-layer interactions, i.e. the hole continuous to propagate in the same layer after the interaction. The red vertex is associated with  interactions where the hole jumps from one layer to the other. The propagators containing two colors represent $G_{\text{o}}$.
	}
	\label{Fig:Self}
\end{figure}

To proceed with the calculation of these self-energies, we now invoke our essential approximation scheme, i.e. the self-consistent Born approximation (SCBA). Diagrammatically, the SCBA corresponds to the inclusion of all rainbow diagrams, as shown in Fig. \ref{Fig:Self}. 
  By using the symmetries in Eq.~\eqref{Eq:interaction_vertex_symmetry} and the expressions for the Green's functions Eq. \eqref{Eq:Prop_D_O}, the self-energies can for zero temperature 
  be written in a compact matrix form as 
 \begin{align}
\Sigma(\bp,\omega) = 2 \sum_{\bk} V^2_{+}(\bp,\bk) G(\bp + \bk,\omega - \omega_{+,\bk}),
\label{Eq:Self_Energies}
\end{align}
where 
\begin{align}
V_{+}(\bp,\bk) = \begin{bmatrix} g_{+}(\bp,\bk) & f_{+}(\bk) \\ f_{+}(\bk) & g_{+}(\bp,\bk) \end{bmatrix},
\label{Eq:Coupling_Matrix}
\end{align}
and the factor of $2$ reflects that the two spin wave branches contribute equally to the self-energies. Here, we have performed the usual analytical continuation $i\omega_p \to \omega + i0_+$ to get the retarded Green's functions. The structure of Eq.~\eqref{Eq:Self_Energies} 
 for the self-energy is the same as for a single layer \cite{kane1989,martinez1991} with the vertex functions replaced by matrices. 
  Equations \eqref{Eq:Prop_D_O}-\eqref{Eq:Self_Energies} constitute our self-consistent equations, which we solve iteratively starting from $\Sigma= 0$.

\section{Equilibrium properties}\label{Sec:Spectral}
We now analyze the equilibrium properties of  mobile holes in the bilayer. While this problem has been explored by several authors, we focus on aspects important for the new generation of optical lattice experiments. The analysis in this section also lays the foundation for the understanding non-equilibrium dynamics described in Sec.~\ref{NonequilSec}. 

We calculate the hole spectral functions ${A}_{\rm d/\rm o}(\bp,\omega)= -2\text{Im}[G_{\rm d/\rm o}(\bp,\omega)]$ by 
  solving \eqref{Eq:Prop_D_O}-\eqref{Eq:Self_Energies} numerically for a pair of $28\times 28$ lattices.
The diagonal spectral function ${A}_{\rm d}(\bp,\omega)$ is associated with the motion of a hole within a given layer whereas the off-diagonal spectral 
function ${A}_{\rm o}(\bp,\omega)$ is associated with the hole starting in one layer and ending up in the other. 
Using the interlayer hopping matrix element $t$ as an energy unit, we have three free parameters: $J/t$, $J_\perp/t$, and $t_\perp/t$.
To be specific and to reduce parameter space, we vary $J/t$ and $t_\perp/t$ keeping $J_\perp =4t_\perp^2/U=(t_{\perp}/t)^{2} J$  inspired by the connection 
 to the Hubbard model.

\begin{figure}
	\hspace{-0.5cm}
	\includegraphics[width=\columnwidth]{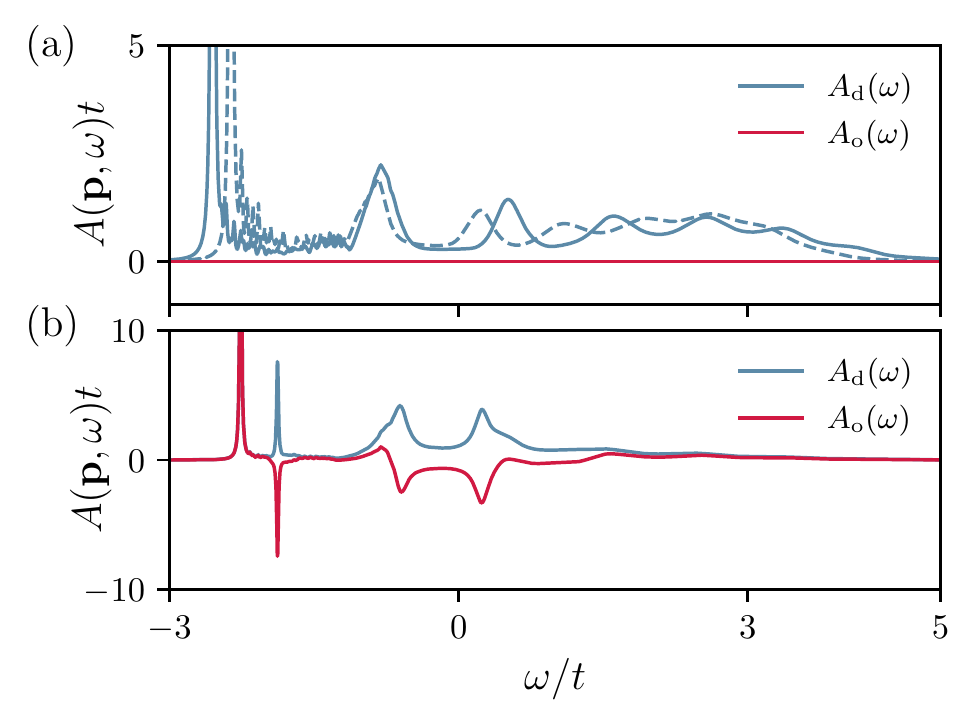}
        \caption{{\bf Hole spectral functions.} Panel $\mathrm{(a)}$ shows the spectral functions for $\bp=(\pi/2,\pi/2)$
        with $J/t=0.3$ and  $t_{\perp}/t=0$ (dashed) as well as $t_{\perp}/t=1$ (solid). The quasiparticle peaks at $\omega/t \simeq -2.3$ and $\omega/t \simeq -2.5$
        for $t_{\perp}/t=0$ and $t_{\perp}/t=1$ respectively give the energy of the magnetic polaron. As expected, the off-diagonal spectral function vanishes for  
         $t_{\perp}=0$ and it also vanishes for $t_{\perp}/t=1$ due to the AF order. 
        Panel $\mathrm{(b)}$ shows the spectral functions for $\bp=(0,0)$ where both the diagonal and off-diagonal parts are non-zero  and have the same poles.}
        \label{Fig:SpectralFns}
\end{figure}

Figure \ref{Fig:SpectralFns}(a) shows the hole spectral functions for the momentum $\bp =\bQ/2= (\pi/2,\pi/2)$,  $J/t = 0.3$, and two different values of $t_\perp/t$.  
Consider first the case of a vanishing interlayer hopping $t_{\perp}/t = 0$, which corresponds to the two layers being  decoupled  so that 
${A}_{\rm o}=0$. 
The diagonal spectral function ${A}_{\rm d}$  exhibits a clear quasiparticle peak at $\omega/t \simeq -2.3$ in agreement with previous results for a single layer magnetic polaron~\cite{nielsen2021}. This polaron is one of four degenerate ground states with momenta $\bp=(\pm \pi/2, \pm \pi/2)$ 
for each layer. 
 The broad peaks at higher energies in Fig.~\ref{Fig:SpectralFns}(a)
  can be interpreted as damped string excitations of the magnetic polaron~\cite{nielsen2022}.

Consider next the case $t_{\perp}/t = 1$. The diagonal spectral function ${A}_{\rm d}$ in 
Fig.~\ref{Fig:SpectralFns}(a) shows that the energy of the magnetic polaron is now lowered to $\omega/t \simeq -2.5$ due to the coupling between the two layers. 
Remarkably, the off-diagonal spectral function remains strictly zero even though there is now tunneling between the two layers. 
To understand this, one can use the symmetries in  Eq.~\eqref{Eq:interaction_vertex_symmetry} together with $\omega_{+,\bk+\mathbf{Q}} = \omega_{-,\bk}$ to show
\begin{align}
	G_{\rm{d}}(\mathbf{p}+\mathbf{Q},\omega) &= G_{\rm{d}}(\mathbf{p},\omega) \nonumber\\
	 G_{\rm{o}}(\mathbf{p}+\mathbf{Q},\omega) &= -G_{\rm{o}}(\mathbf{p},\omega).
	\label{Eq:self_greens_symmetry}
\end{align}
It follows from Eq.~\eqref{Eq:self_greens_symmetry} together with the inversion symmetry that indeed $G_{\rm{o}}(\mathbf{Q}/2,\omega)=0$. 
In fact, one can show that  $G_{\rm o}({\bf{p}},\omega)=0$ for all momenta along the edges of the magnetic BZ defined by $|k_x|+|k_y|=\pi$~\cite{yin1998}. 
The symmetry given by Eq.~\eqref{Eq:self_greens_symmetry} can also be inferred from a  semi-classical picture. Assume the hole is initially created in sublattice $\rm{A}$
 in layer 1. It will then create magnetic frustration, i.e.\ aligned spins, when it jumps to layer 2. As can be seen in Fig. \ref{Fig:Bilayer}, 
 these aligned spins can be repaired by the $\hat{S}^{+}\hat{S}^{-}$ terms  in Eq.~\eqref{Eq:HJ} only when the hole resides in sublattice $\rm{A}$ in layer 2, i.e.\ when 
 it has performed an even number of jumps. 
 Hence, $\rG_{\rm o}({\bf r} \in B,\omega)=0$ which in momentum space translates to 
 \begin{align}
	\rG_{\rm o}(\bp + \bQ,\omega) =\sum_{\br \in {\rm A}} \frac{e^{i \bQ \cdot \br }}{\sqrt N} e^{i \bp \cdot \br }  \rG_{\rm o}(\br,\omega)
	 = - \rG_{\rm o}(\bp,\omega),
	 \label{Eq:AntiSym}
\end{align} 
where $\br \in {\rm A}$ indicates sublattice A in the opposite  layer of where the hole was created. 

From $G_{\rm o}({\mathbf Q}/2,\omega)=0$ it follows that the splitting between the symmetric and anti-symmetric 
polarons in Eq.~\eqref{Eq:SymAntiSymWavefn} vanish and there are two degenerate polaron states. Thus, one can for momentum ${\mathbf Q}/2$ rotate to polaron eigenstates where the 
bare hole is exclusively in one layer, say $1$, so that $|\Psi_{\mathbf p}\rangle=\sqrt{Z_{\mathbf p}}\hat h^\dagger_{1,\mathbf p}|\text{AF}\rangle+\ldots$, 
 irrespective of the value of $t_{\perp}/t$. For $t_{\perp}/t>0$, the hole can of course jump into the other layer but this will always result in spin waves in the system. 

\begin{figure}[!t]
 \includegraphics[width=\columnwidth]{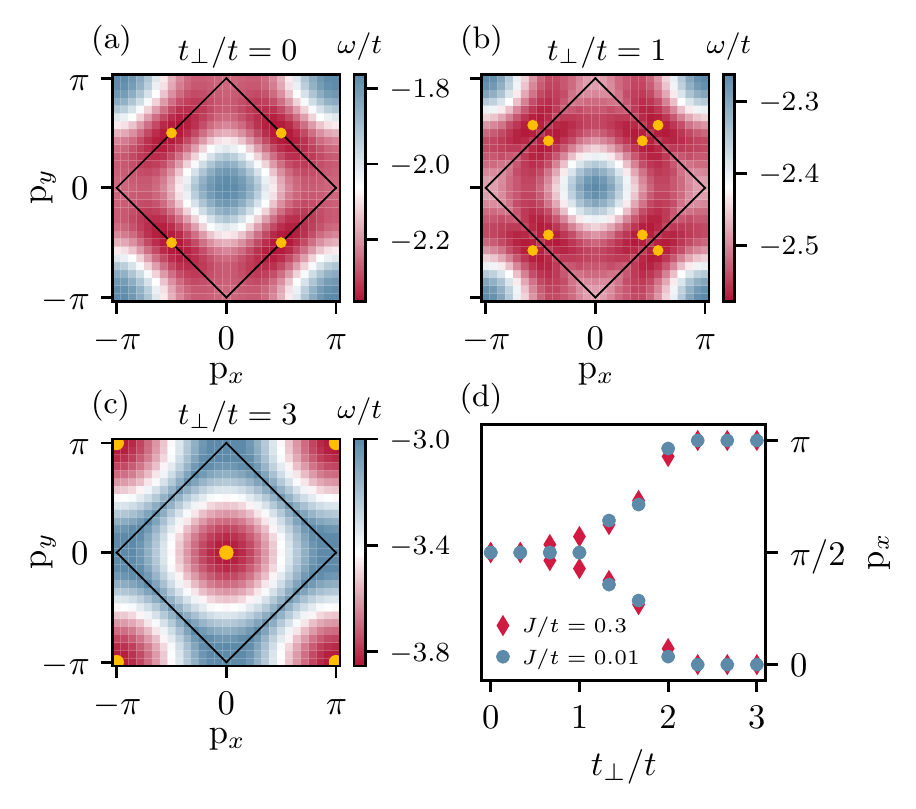}
 \caption { {\bf Magnetic polaron dispersion.} Panel (a)-(c) show the energy of the magnetic polaron in the BZ  for 
 $J/t=0.3$ and $t_{\perp}/t = 0,1,3$. The yellow dots indicate the minima and the black square  the edge of the magnetic BZ given by 
  $|{\mathrm p}_x|+|{\mathrm p}_y|=\pi$. Panel (d) shows the momentum $\mathbf{p}=(\mathrm{p}_x,\mathrm{p}_x)$ of the ground state as a function of $t_{\perp}/t$ for $J/t=0.01$ and $J/t=0.3$. Note that the mirror symmetry around the boundary of the magnetic BZ gives rise to several degenerate minima of the polaron energy. } 
  \label{Fig:EnergyinBZ}
\end{figure}

The degeneracy of the symmetric and antisymmetic polarons is broken when the momentum is not on the edge of the magnetic BZ and $t_\perp\neq 0$.
 This can be seen in 
Fig.~\ref{Fig:SpectralFns}(b), which shows  the spectral functions for  $\bp = \bf{0}$,  $J/t =0.3$, and $t_\perp/t =1$. The   
off-diagonal spectral function ${A}_{\rm o}$  is now non-zero due to the coupling between the two layers. Equivalently, there is an energy splitting between
 the symmetric and anti-symmetric polarons, and 
 the  spectral functions have two peaks at the polaron energies $\omega=\epsilon^{\pm}_{\mathbf p}$ with strengths $Z^{\pm}_{\mathbf p}$
for  ${A}_{\rm d}$ and  $\pm Z^{\pm}_{\mathbf p}$ for ${A}_{\rm o}$ as can  be seen in Fig. \ref{Fig:SpectralFns}(b). 
As a consistency and accuracy check, we find that the numerics reproduce the sum rule $\int_{-\infty}^{\infty}{A}_{\rm d}({\mathbf p},\omega)=2\pi$ with a deviation less than $ 1\% $ and 
$\int_{-\infty}^{\infty}{A}_{\rm o}({\mathbf p},\omega)=0$ with a deviation $|\int_{-\infty}^{\infty}{A}_{\rm o}({\mathbf p},\omega)| < 0.01$, see Appendix \ref{app:SumRule} for details. In Fig.~\ref{fig:contour}, we plot the spectral functions  as a function of frequency $\omega$  along the diagonal $\bp = {(\mathrm{p}_x,\mathrm{p}_x)}$ in the BZ for $J/t = 0.3$ and different values of the interlayer hopping $t_\perp$. These plots clearly show the two quasiparticle  branches corresponding to symmetric and anti-symmetric  polarons, which are degenerate 
at ${\mathbf p}={\mathbf Q}/2$. 
We also see how the momentum of the ground state becomes different from ${\mathbf Q}/2$ with increasing interlayer hopping. 
This is illustrated further in Fig.~\ref{Fig:EnergyinBZ}(a)-(c), which show the polaron spectrum in the 
BZ for $J/t = 0.3$ and different values of $t_\perp$. The momentum of the ground state indicated by yellow dots 
gradually move away from ${\mathbf Q}/2$ with increasing interlayer hopping. 
The mirror symmetry around the boundary of the magnetic BZ then gives rise to eight minima of the polaron energy. 
Eventually, the minimum settles at $\mathbf{ p = 0}$ (and $\mathbf{Q}$).  
The momentum of the ground state is plotted in Fig.~\ref{Fig:EnergyinBZ}(d) as function of $t_\perp/t$ showing that 
it moves away from $\mathbf{p = Q}/2$ for $t_{\perp}/t \approx 0.3$ for $J/t=0.3$ and $J/t=1$ for  $J/t=0.01$. Such
a change of the ground state momentum   from ${\mathbf Q}/2$ to $\mathbf{ 0}$  was also predicted using a variational approach~\cite{vojta1999}.

\begin{figure}[!t]
	\hspace{-0.35cm}
	\includegraphics[width=\columnwidth]{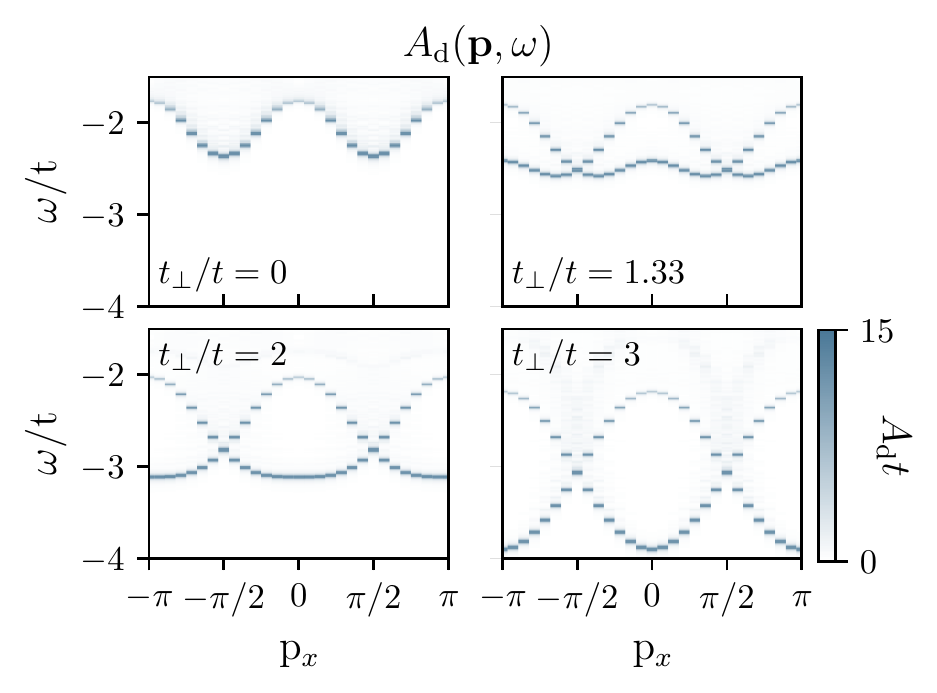}
	\caption{{\bf Polaron energy bands.} The polaron energy bands along the diagonal $\bp = (\mathrm{p}_x,\mathrm{p}_x)$ for $J/t = 0.3$ and different values of $t_\perp$. This shows the splitting of the symmetric and anti-symmetric polaron energies when $t_{\perp}\neq0$ except for $\mathbf{p} = \mathbf{Q}/2$ where the two states are degenerate. The momentum of the ground state also moves from $ \mathbf{Q}$ to ${\mathbf 0}$ with increasing $t_{\perp}$.}
	\label{fig:contour}
\end{figure}

\section{Non-equilibrium  dynamics}\label{NonequilSec}
Having explored the fundamental equilibrium properties of a hole in the AF bilayer, we now turn to the non-equilibrium dynamics,  
which can be probed with unprecedented resolution in a new generation of optical lattice experiments~\cite{Christie2019,Brown2019,Koepsell:2019ua,Ji2021,Koepsell2021,gall2021a,Hirthe2022}.

We imagine a hole created at a given lattice site and analyze its subsequent dynamics. 
The key object to calculate for this kind of experiment is 
the Green's function $G_{lm}^{>}({\mathbf p},\tau)=-i\langle \hat{h}_{l,{\mathbf p}}(\tau)\hat{h}^{\dagger}_{m,{\mathbf p}}(0)\rangle$, 
which gives the overlap between an initial state corresponding to a hole with momentum ${\mathbf p}$ in  layer $m$ at time $0$, and a state where the 
hole is in layer  $l$ at time $\tau$, which should not be confused with the imaginary time. This, hereby, gives access to the full dynamics of the lowest order coefficient in the many-body wave function~\cite{nielsen2022}.
While one in general needs a formalism such as Keldysh Green's functions to calculate  non-equilibrium many-body physics, it turns out that 
we can obtain the hole dynamics from our  SCBA calculation by analytic continuation. The reason is that for a single hole, we have 
$iG^>_{ll}(\bp,\omega)={A}_\rd(\bp,\omega)$ and $iG^>_{l\neq m}(\bp,\omega)={A}_\ro(\bp,\omega)$~\cite{skou2021,bruus2004}. 
Hence, we can  obtain the real time dynamics by Fourier transforming the spectral functions $\mathrm{A_{\rd/\ro}}(\mathbf{p},\omega)$.
 
 \subsection{Interlayer oscillations}  
As we saw above, the spectral function in general consists of two quasiparticle peaks and a many-body continuum. 
In analogy with what has been observed for a mobile impurity atoms in  atomic gases~\cite{Nielsen_2019,skou2021}, the continuum eventually
 decoheres so that the long time dynamics is governed by polaron formation. As a result, the real-time Green's function approaches
\begin{align}
	iG^>_{\text{d}/\text{o}}(\bp,\tau) \to \frac{1}{2}\left( Z^{+}_{\bp } e^{ -i\varepsilon^{+}_{ \bp }\tau } \pm Z^{-}_{\bp } e^{ -i\varepsilon^{-}_{ \bp }\tau} \right)
\end{align}
for long times, $\tau \gg 1/t$. Hence,
\begin{align}
|G^>_{\text{d}}(\bp,\tau)| &=  \frac{Z^{+}_{\bp }}{2} \sqrt{ 1 + \left( \frac{Z^{-}_{ \bp } }{Z^{+}_{ \bp } }\right)^{2} + 2\frac{Z^{-}_{\bp }}{Z^{+}_{\bp }} \cos{( \Delta \varepsilon_{\bp} \tau )} } \nonumber  \\
	|G^>_{\text{o}}(\bp,\tau)| &= \frac{Z^{+}_{\bp }}{2}   \sqrt{ 1 + \left( \frac{Z^{-}_{ \bp } }{Z^{+}_{ \bp } }\right)^{2} - 2\frac{Z^{-}_{\bp }}{Z^{+}_{\bp }} \cos{( \Delta \varepsilon_{\bp} \tau )} }, \label{Eq:A_time}
\end{align} 
with $\Delta \varepsilon_{\bp} = \varepsilon^{+}_{\bp} - \varepsilon^{-}_{\bp}$ the energy difference between the symmetric and anti-symmetric polaron. 

\begin{figure}[!t]
        \centering
        \begin{subfigure}[!hb]{0.45\textwidth}
        		\centering
        		\includegraphics[width=\textwidth]{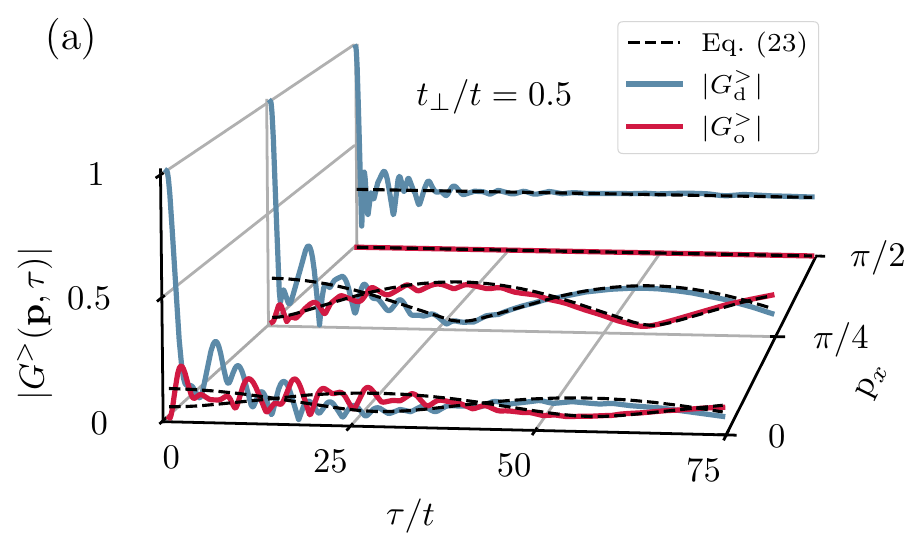}
        \end{subfigure}
        \begin{subfigure}[!hb]{0.45\textwidth}
		\centering
		\includegraphics[width=\textwidth]{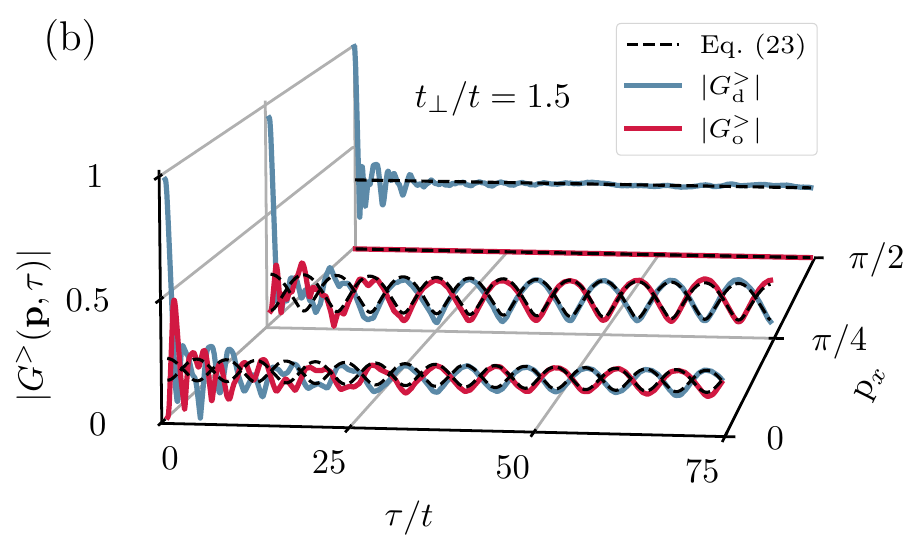}
        \end{subfigure}
        \caption {{\bf Hole oscillating between the layers.} The diagonal and off-diagonal hole Green's functions $G^>_{\rm d/o}(\bp,\tau)$as a function of time for different momenta  ${\mathbf p}=(\rp_x,\rp_x)$. The long time prediction given by Eq.~\eqref{Eq:A_time} is plotted as a dashed line. 
      } 
        \label{fig:time}
\end{figure}

In Fig.~\ref{fig:time}, we plot the hole dynamics as described by $|G^>_{\text{d}}(\bp,\tau)|$ and $|G^>_{\text{o}}(\bp,\tau)|$ for different momenta 
${\mathbf p}={(\mathrm{p}_x,\mathrm{p}_x)}$ along the BZ diagonal, taking $J/t=0.3$, $t_\perp/t=0.5$(a) and $t_\perp/t=1.5$(b).  
 Initially, $|G^>_{\text{d}}(\bp,\tau)|$ decreases from unity simply reflecting that the hole starts to generate spin waves.  
 On the other hand, $|G^>_{\text{o}}(\bp,\tau)|$ increases from zero because the hole created in a given layer jumps to the other layer, except for ${\mathbf p}=(\pi/2,\pi/2)$ 
 where it remains zero.  The short time dynamics is, therefore, faster for  $t_\perp/t=1.5$ than for  $t_\perp/t=0.5$ as can be seen by comparing 
 Figs.~\ref{fig:time}(a) and (b). At later times,  the population of the  hole is clearly seen to  oscillate between the two layers, which is very 
  accurately described by Eq.~\eqref{Eq:A_time}, confirming that the many-body continuum indeed decoheres for long times $\tau\gg t$. 
Physically, these oscillations arise from the fact that a hole initially localized in one layer corresponds to an equal superposition of the symmetric and anti-symmetric polaron. As time evolves, this results in  a beating with a frequency given by their energy difference. The oscillations are faster for $t_\perp/t=1.5$ than for  $t_\perp/t=0.5$, since a stronger coupling between the planes results in a larger energy splitting between the two polarons. There are, however, no oscillations for ${\mathbf p}={\mathbf Q}/2$ where the symmetric and anti-symmetric polarons are degenerate and the bare hole somewhat counterintuitively remains in one layer for all times. As discussed above, this is because the symmetric and anti-symmetric polarons are degenerate for ${\mathbf p}={\mathbf Q}/2$ so 
that a polaron eigenstate can be formed with the bare hole exclusively in one layer,  while its presence in the other layer is always accompanied by spin waves. The excellent agreement between Eq.~\eqref{Eq:A_time} and the numerics also illustrates that  the dynamics of the bare hole and the polaron is strongly entangled, since Eq. \eqref{Eq:A_time} arises from only considering the polaron states whereas the Green's functions give the dynamics of the bare hole.

Experimentally, this intriguing interlayer dynamics of the hole is most easily measured for the ${\mathbf p}=0$ case, since it corresponds to the hole initially being created with uniform density and no relative phase over one layer. The excited spin waves will change these oscillations, but using the form of the wave function in Eq. \eqref{Eq:SymAntiSymWavefn} shows that the oscillation frequency persists.

\subsection{Site resolved dynamics}
In a recent optical lattice experiment, the motion of a hole initially created at a given lattice site was observed with single site resolution in a one AF layer~\cite{Ji2021}.
Inspired by this impressive experiment, we now explore the hole dynamics in real space via the Green's functions $G^>_{\text{d}}(\br,\tau)$ and $G^>_{\text{o}}(\br,\tau)$ obtained by Fourier transforming from momentum space. They give the overlap between an initial state with a hole created at the origin at time $0$, and a state where the hole is removed in the same/different layer at position $\br$ and time $\tau$. 

\begin{figure}
	\includegraphics[]{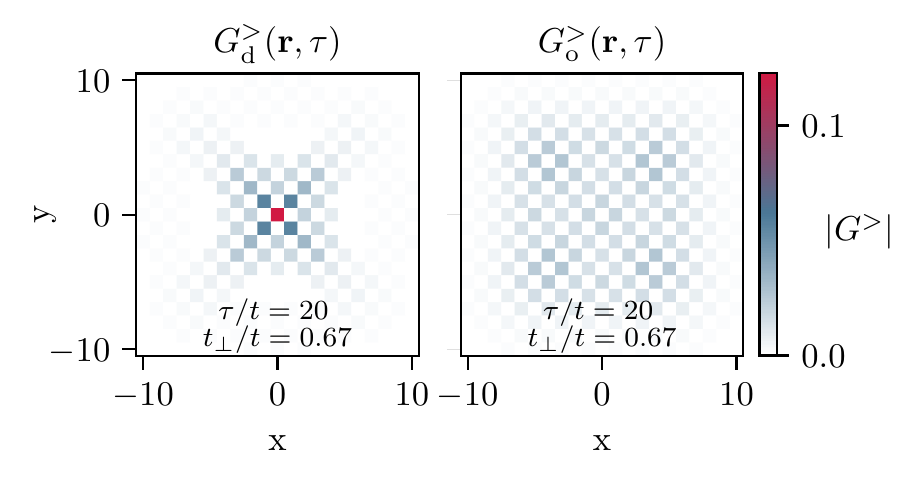}
	\caption{ {\bf Real space hole dynamics.} The diagonal and off-diagonal hole Green's functions in real space for 
	$t_{\perp}/t=0.67$, $J/t=0.3$ at time $\tau/t = 20$. 
	} 
	\label{Fig:RealSpaceSingle}
\end{figure}

In Fig.~\ref{Fig:RealSpaceSingle} we plot $|G^>_{\text{d}}(\br,\tau)|$ and $|G^>_{\text{o}}(\br,\tau)|$  for $J/t=0.3$ $t_\perp/t= 0.67$, and $\tau/t = 20$. 
One clearly sees how the hole has spread out\kristian{,} creating checkerboard patterns. These patterns 
are caused by the fact that the hole must jump an even number of times before the  $\hat{ {S} }^+_{l,\bf i}\hat{ {S} }^-_{l,\bf j}$
terms in Eq.~\eqref{Eq:HJ} can repair the magnetic frustration as discussed in Sec.~\ref{Sec:Spectral}, 
and they are, therefore, inverted with respect to each other in the two layers. Figure \ref{Fig:RealSpaceSingle}, furthermore, shows that the expansion of the  hole is faster along the diagonal directions. This is because the hole motion for long times is determined by the ballistic expansion of magnetic polarons~\cite{nielsen2022} as we will discuss further in Sec.~\ref{phasetrans}. Since the polaron dispersion is steepest along the diagonals as can be seen in Fig.~\ref{Fig:EnergyinBZ}, this results in a faster expansion along those directions. 
 
We remind the reader that  $G^{>}_{\text{d}/\text{o}}(\bp,\tau)$ describes the dynamics of a \emph{bare} hole in the sense that it gives the overlap between hole states separated by the time $\tau$ with no spin waves present. Hence, it does not give information regarding the overlap with final states describing the simultaneous presence of a hole and spin waves such as those connected to the origin by an odd number of jumps.  As such, the checkerboard patterns shown in Fig.~\ref{Fig:RealSpaceSingle} is an artefact of projecting out these states. Nevertheless, the Green's functions reflect the hole dynamics, since a bare hole is strongly entangled with the polarons. We discuss this point further in the next section. 

\subsection{Order to disorder quantum phase transition}\label{phasetrans}
It is well-known that in the absence  of a hole, the bilayer system undergoes a quantum phase transition  with increasing $t_\perp/t$ 
from the ordered AF state to a disordered  state, where neighbouring spins in the two layers form spin singlets~\cite{chubukov1995, scalettar1994, millis1994}. Quantum Monte-Carlo calculations and series expansions  yield the critical value $J_\perp/J\sim 2.5$ for this transition~\cite{Hida1992,Sandvik1994,Sandvik1995}, corresponding to $t_\perp/t=1.58$ when $J_\perp/J=t_\perp^2/t^2$.  The presence of  a single hole will not affect this phase transition, whereas the phase transition does affect the hole dynamics as we will now explore.  

\begin{figure}
	\centering
	\includegraphics[width=\columnwidth]{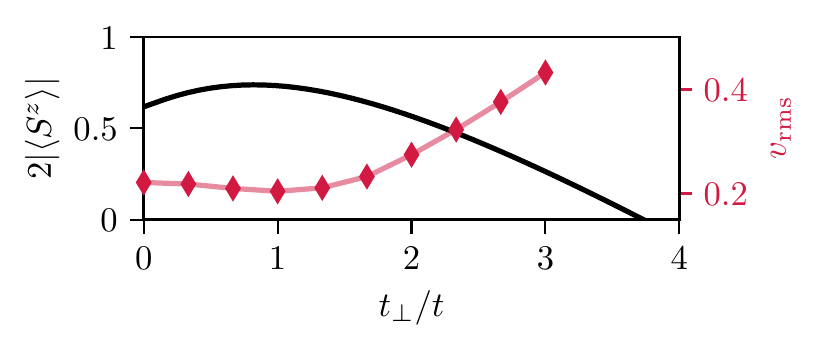}
	\caption{{\bf Magnetization and hole  velocity.} The sublattice magnetization given by Eq. \eqref{Eq:Corr} (black),  and the final rms velocity of the hole (red). The magnetization shown is scaled such that unity corresponds to perfect AF alignment.} 
	\label{fig:SpinCorr}
\end{figure}

In Fig.~\ref{fig:SpinCorr}, we plot the sublattice magnetization as a function of $t_\perp/t$ calculated within linear spin wave theory as 
\begin{align}
	|\langle {S}^{z} \rangle| = \frac{1}{2}-\frac{1}{2N} \sum_{\mathbf{k}} \left( v_{\mathbf{k},+}^{2} + v_{\mathbf{k},-}^{2} \right). 
	\label{Eq:Corr}
\end{align}
The magnetization first increases with $t_{\perp}/t$ reaching a maximum at $t_{\perp}/t = 0.82$ after which it decreases until magnetic order is lost at $t_{\perp}/t = 3.67$ in agreement with Ref.~\cite{chubukov1995}. Linear spin wave theory thus captures the qualitative physics of the phase transition with the AF order going to zero,  but as usual for a mean-field theory, it overestimates 
the transition point due to the omission of longitudinal  spin fluctuations~\cite{chubukov1995}. 

\begin{figure}[!t]
	\centering
	\includegraphics{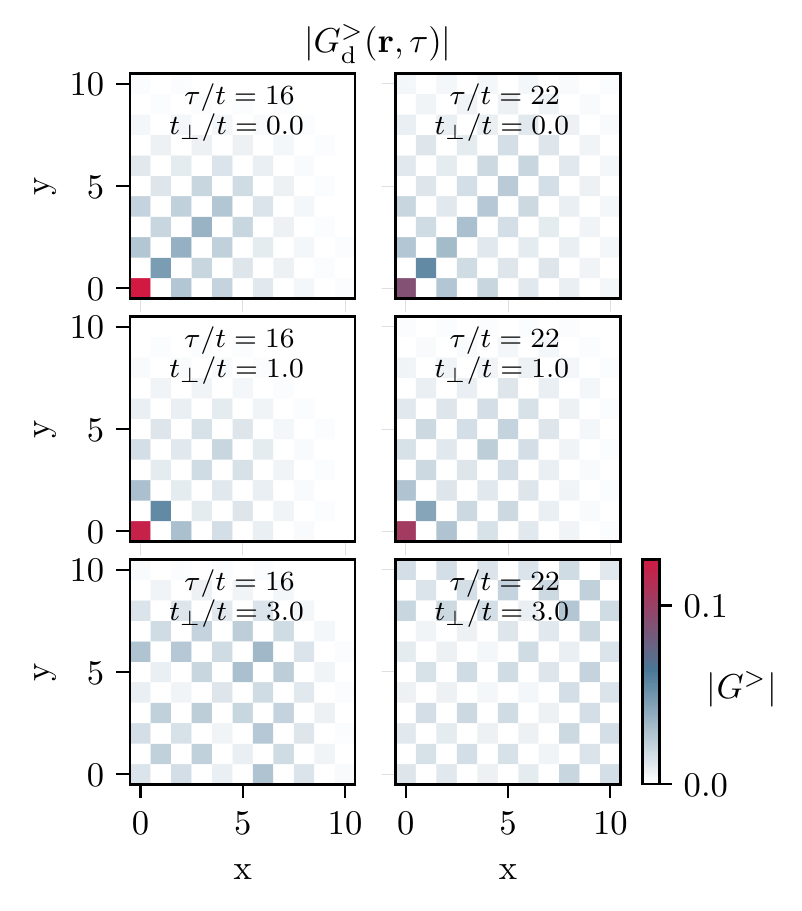}
	\caption{ {\bf Hole expansion for different interlayer couplings.} The real space diagonal hole Green's function for $J/t=0.3$, $t_{\perp}/t=0,1,3$, and $\tau/t = 15,22$. }
	\label{Fig:RealSpaceTriple}
\end{figure}

To explore the effects of this phase transition on the hole dynamics, we plot in Fig.~\ref{Fig:RealSpaceTriple}  the diagonal and off-diagonal real space Green's functions for $\tau/t=16$ and $\tau/t=22$. This shows that the hole delocalizes slower for $t_{\perp}/t =1$  compared to $t_{\perp}/t = 0$. Physically, this is because the AF order is larger at $t_{\perp}/t =1$, see Fig.~\ref{fig:SpinCorr}, making it energetically more costly for the hole to move. Increasing the interlayer coupling further, we see that the hole delocalizes faster for $t_{\perp}/t =3$. This non-monotonic behaviour of the hole expansion velocity is consistent with the magnetic order, which first increases with increasing interlayer coupling before it vanishes at the phase transition.

 To quantify this,  we consider the rms distance of the bare hole from origin defined by
\begin{align}
	d_{\mathrm{rms}}(\tau) = \sqrt{\frac{\sum_{{\bf r}} r^{2}\left( | G^>_{\rm d}({\bf r},\tau)  |^{2}+ | G^>_{\rm o}({\bf r},\tau)  |^{2} \right) }{\sum_{{\bf r}} \left(|G^>_{\rm d}({\bf r},\tau)  |^{2} + | G^>_{\rm o}({\bf r},\tau)  |^{2} \right) }}.
	\label{Eq:rms}
\end{align}
The normalization takes care of the fact that the likelihood of finding a bare hole is less than unity for $\tau>0$ where spin waves are present. Indeed, we have $\sum_{{\bf r}} (| G^>_{\rm d}({\bf r},\tau)  |^{2} + | G^>_{\rm o}({\bf r},\tau)  |^{2})\rightarrow \sum_{\bf{p}}( ({ Z^{+}_{\bf{p}}})^{2} + ({Z^{-}_{\bf{p}}})^{2})/2N$
for $\tau/t\rightarrow \infty$ when the  many-body continuum has decohered.

\begin{figure}[!t]
\includegraphics[width=\columnwidth]{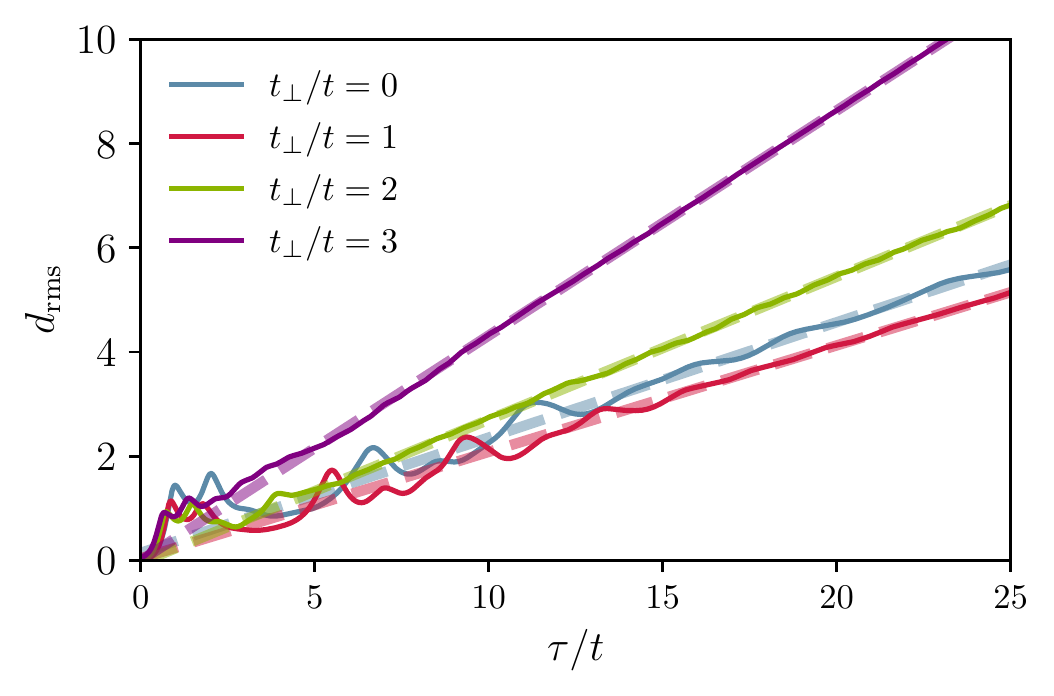}
\caption{ {\bf rms distance of the hole.} The rms distance given by Eq.~\ref{Eq:rms}  for $J/t=0.3$ and $t_{\perp}/t = 0,1,2,3$. 
The final expansion velocities are in Fig. \ref{fig:SpinCorr} are found from the slopes of the asymptotic dashed lines. } 
\label{fig:RMSdistance}
\end{figure}

In Fig.~\ref{fig:RMSdistance}, we plot  $d_{\mathrm{rms}}(\tau)$  for $J/t=0.3$ and $t_{\perp}/t = 0,1,2,3$.
After an initial rapid expansion and oscillations, the hole starts moving with a constant velocity. This 
 long time dynamics of the hole  is governed by  ballistic expansion of magnetic polarons, which have been formed after the 
initial creation of a bare hole. Indeed, it has recently been shown using a time-dependent wave function within the SCBA that the long time expansion of a hole initially 
created at a given site in a single layer is determined by the formation and ballistic motion of magnetic polarons~\cite{nielsen2022}. 
Here, we find  $v_{\text{rms}}=\dot d_\text{rms}=0.213 t$ for the expansion  velocity of the bare hole at long times in the case of the two layers completely decoupled with 
  $t_{\perp}/t = 0$. This is very close to value $v_{\mathrm{rms}}=0.215 t$ found for the 
 final expansion velocity of  magnetic polarons for a single layer~\cite{nielsen2022}. The agreement  explicitly demonstrates that the motion of the bare hole discussed in this 
 paper closely follows the motion of magnetic polarons for long times, which can be understood from the fact that the hole is closely entangled with the polaron for 
 long times. 
 
Figure \ref{fig:RMSdistance}, furthermore, shows that the hole expands slower for $t_\perp/t=1$ as compared to $t_\perp/t=0$, and that the expansion velocity then increases again 
with $t_\perp/t=3$, consistent with Fig.~\ref{Fig:RealSpaceTriple}. This is  illustrated further in Fig.~\ref{fig:SpinCorr} where the final expansion velocity after the magnetic polaron has formed is plotted 
as a function of $t_{\perp}/t$. The expansion velocity depends non-monotonically 
 on $t_{\perp}/t = 0$, directly reflecting the behaviour of the magnetic order also shown in Fig.~\ref{fig:SpinCorr}. 
 From this  we conclude that the approach to the order-to-disorder phase transition of the bilayer system with increasing interlayer coupling can be detected 
 as an increase in the hole expansion velocity as the transition point is approached from the AF phase,  reflecting the decrease in the magnetic energy cost of hole hopping. 

It should be noted that while our theory does not contain the singlet correlations across the layers giving rise to the disordered phase, we believe 
our results are qualitatively reliable. In particular, the increase in the expansion velocity as the phase transition is approached inside the AF phase is a
robust result, since a decreasing magnetic order means less energy cost for hole hopping, which physically must be expected to lead to a higher velocity. 
Also, our theory agrees qualitatively with a variational calculation interpolating between the ordered AF state and a disordered singlet product state regarding the influence of the interlayer coupling on the polaron spectrum~\cite{vojta1999}. In this paper, the momentum of the ground state was also predicted to move from ${\mathbf Q}/2$ to  ${\mathbf 0}$ with increasing interlayer coupling $J_\perp$ in agreement with our findings shown in Fig.~\ref{Fig:EnergyinBZ}. Band structures similar to what we show in Fig.~\ref{fig:contour} were also reported. So even though our analysis for the phase transition is quantitatively unreliable, we expect the main finding, i.e.\ the speed up of the hole, to be qualitatively correct. 

\section{Conclusions}\label{Conclusions and outlook}
We investigated the equilibrium and non-equilibrium properties of a hole in an AF bilayer using a diagrammatic approach based on the SCBA. The spectral properties of the hole were shown 
to exhibit two quasiparticle peaks corresponding to magnetic polarons that are either symmetric or anti-symmetric under layer exchange. 
We calculated the energy bands of these two kinds of polarons in the BZ and showed that they are degenerate at certain momenta due to the underlying AF symmetry. The momentum of the 
ground state polaron was, furthermore, shown to move to the origin of the BZ with increasing interlayer coupling. We then demonstrated that  a hole initially created in one layer, 
oscillates between the two layers with a  frequency determined by the energy difference between the symmetric and the anti-symmetric polaron. Finally, 
we analyzed how the asymptotic expansion velocity of a hole initially created at a given lattice site is governed by the ballistic motion of polarons, and that it first decreases 
and then increases as a function of the interlayer coupling, reflecting that a 
quantum phase transition to a disordered state of the spins is approached. 

We have largely focuzed on observables that are accessible in experiments based on cold atoms in optical lattices. In particular, their single site resolution  
and the ability to perform quench experiments where a hole is abruptly released from a given lattice site~\cite{Christie2019,Brown2019,Koepsell:2019ua,Ji2021,Koepsell2021}, 
combined with  the recent creation of  a bilayer system~\cite{gall2021a}, show that optical lattices   provide a promising platform 
 to observe the effects discussed above. 

Interesting future research directions include calculating the  time-dependent wave function of a hole initially created at a given lattice in the bilayer using the SCBA. This has recently been achieved for a single layer~\cite{nielsen2022} and would give access to the full non-equilibrium dynamics of the hole. Another intriguing question concerns the 
properties of a hole in the disordered phase for large interlayer coupling, where neighbouring spins in the two layers form singlets. Finally, an important problem concerns the role of 
temperature, since optical lattice experiments invariably are performed at a non-zero temperature. 

{\it Acknowledgments.-} 
This work has been supported by the Danish National Research Foundation through the Center of Excellence "CCQ" (Grant agreement no.: DNRF156), as well as the Carlsberg Foundation through a Carlsberg Internationalisation Fellowship. We thank T. Pohl for useful comments and discussions.



\appendix

\section{Diagonalization of $\hat{H}_{J}$} \label{app:Diagonalization}
In this appendix, we will elaborate on the steps taken in diagonalizing $\hat{H}_{J}$, Eq. \eqref{Eq:HJ}. Using the mappings described in Section \ref{sec:model} and neglecting the hardcore constraint we find
\begin{align}
	\hat{H}_{J} = & \ \hat{H}^{int}_J +
	\frac{J}{2}\sum_{\langle \bi \bj \rangle,l} \Big[ \hat{s}^{\dagger}_{l,\bi}\hat{s}_{l,\bi} + \hat{s}^{\dagger}_{l,\bj}\hat{s}_{l,\bj} \nonumber \\  
	&+ \alpha \left( \hat{s}_{l,\bi}\hat{s}_{l,\bj} + \hat{s}^{\dagger}_{l,\bj}\hat{s}^{\dagger}_{l,\bi}  \right) \Big]  - \frac{Jz}{2}\left( N -1\right) \nonumber \\
	&+\frac{J_{\perp}}{2}\sum_{i} \Big[ \hat{s}^{\dagger}_{l,\bi}\hat{s}_{l,\bi} + \hat{s}^{\dagger}_{\bar{l},\bi}\hat{s}_{\bar{l},\bi}   \nonumber \\  
	&+\alpha_{\perp}  \left( \hat{s}_{l,\bi}\hat{s}_{\bar{l},\bi} + \hat{s}^{\dagger}_{\bar{l},\bi}\hat{s}^{\dagger}_{l,\bi}  \right)  \Big]  - \frac{J_{\perp}}{2}\left(N-1\right),
	\label{Eq:app_H_J}
\end{align}
where $\hat{H}^{int}_J$ contains the non-linear terms which are neglected. $\alpha$ and $\alpha_{\perp}$ describe possible intra and inter layer anisotropy respectively. They are set equal to unity in the main text for the sake of clarity, but they can easily be considered. Defining the Fourier transform in the standard way
\begin{align}
	\hat{s}_{l,\bi} = \frac{1}{\sqrt{N}} \sum_{\textbf{k}} e^{i \textbf{k} \cdot \bi } \hat{s}_{l,\textbf{k}},
\end{align}
the Hamiltonian in momentum space can be written as
\begin{align}
	&\hat{H}^{int}_J  = E_{0} + \frac{1}{4}\sum_{\textbf{k}} 
	\nonumber \\ 
	&\begin{bmatrix}
   		\hat{s}^{\dagger}_{1,\textbf{k}} & \hat{s}_{1,-\textbf{k}}  & \hat{s}^{\dagger}_{2,\textbf{k}} & \hat{s}_{2,-\textbf{k}} 
	\end{bmatrix} 
	\begin{bmatrix}
   		\mathcal{H}_{\rm d} & \mathcal{H}_{\rm o} \\
    		\mathcal{H}_{\rm o} & \mathcal{H}_{\rm d}
	\end{bmatrix}
	\begin{bmatrix}
   		\hat{s}_{1,\textbf{k}} \\ 
		\hat{s}^{\dagger}_{1,-\textbf{k}} \\
   		\hat{s}_{2,\textbf{k}} \\ 
		\hat{s}^{\dagger}_{2,-\textbf{k}} 
	\end{bmatrix} \nonumber \\
	\label{App:H_J_k}
\end{align}
with
\begin{align}
 \mathcal{H}_{\rm d} &= \begin{bmatrix}
   					 Jz+J_{\perp}    & \alpha Jz\gamma_{\textbf{k}}  \\
    					\alpha Jz\gamma_{\textbf{k}}    &   Jz+J_{\perp} 
				  \end{bmatrix} \nonumber \\
\mathcal{H}_{\rm o} &=  \begin{bmatrix}
   					 0 & \alpha_{\perp}  J_{\perp} \\
    					 \alpha_{\perp}  J_{\perp}  & 0 
				   \end{bmatrix}.
\end{align}
To diagonalize a Hamiltoninan of this kind, it is practical to get it on a block diagonal form first by performing a unitary transformation. In the block diagonal form, one can then utilize the canonical Bogoliubov transformation for each block individually. To do so, we realize
\begin{align}
	\frac{1}{2}
	\begin{bmatrix}
   		\mathds{1}_2   &  \mathds{1}_2 \\
    		-\mathds{1}_2   &  \mathds{1}_2 
	\end{bmatrix}
	\begin{bmatrix}
   		\mathcal{H}_{\rm d} & \mathcal{H}_{\rm o} \\
    		\mathcal{H}_{\rm o} & \mathcal{H}_{\rm d}
	\end{bmatrix}
	\begin{bmatrix}
   		\mathds{1}_2   &  -\mathds{1}_2 \\
    		\mathds{1}_2   &  \mathds{1}_2 
	\end{bmatrix} = \nonumber \\
	\begin{bmatrix}
   		\mathcal{H}_{\rm d} +  \mathcal{H}_{\rm o} & 0 \\
    		0 & \mathcal{H}_{\rm d} - \mathcal{H}_{\rm o} 
	\end{bmatrix},
\end{align}
such that by defining
 \begin{align}
		\begin{bmatrix}
   		\hat{ s }_{1,\bk} \\ 
			\hat{ s }^{\dagger}_{1,-\bk} \\
   		\hat{ s }_{2,\bk} \\ 
			\hat{ s }^{\dagger}_{2,-\bk} 
		\end{bmatrix}
		=
		 \frac{1}{\sqrt{2}}
		\begin{bmatrix}
   			\mathds{1}_2   &  -\mathds{1}_2 \\
    			\mathds{1}_2   &  \mathds{1}_2 
		\end{bmatrix}
		\begin{bmatrix}
   			U_{+,\bk} & 0 \\
    			0 &  U_{-,\bk}
		\end{bmatrix}
		\begin{bmatrix}
   		\hat{ b}_{+,\bk} \\ 
			\hat{ b}^{\dagger}_{+,-\bk} \\
   		\hat{ b}_{-,\bk} \\ 
			\hat{ b}^{\dagger}_{-,-\bk} 
		\end{bmatrix}.
	\label{App:trans}
\end{align}
and choosing
\begin{equation}
U_{\pm,\bk} = \begin{bmatrix}
   		u_{\pm,\bk} & -v_{\pm,\bk} \\
    		-v_{\pm,\bk} & u_{\pm,\bk}
	\end{bmatrix}
\end{equation}
with the coherence factors given by
\begin{align}
	u_{\pm,\bk} &= \sqrt{\frac{1}{2}\left(\frac{zJ + J_\perp}{2\omega_{\pm,\bk}} + 1\right)}, \nonumber \\ 
	v_{\pm,\bk} &= {\rm sgn}\left[\alpha z J\gamma_\bk \pm \alpha_\perp J_\perp\right]\sqrt{\frac{1}{2}\left(\frac{zJ + J_\perp}{2\omega_{\pm,\bk}} - 1\right)}
\end{align}
the Hamiltonian becomes diagonal
\begin{align}
	\hat{ H }_{J} \simeq E_{0} + \sum_{\bk,\mu = \pm}\omega_{\mu, \bk} \hat{ b}^{\dagger}_{\mu,\bk}\hat{ b}_{\mu,\bk},		
\end{align}
with the dispersion relations 
\begin{equation}
	\omega_{\pm,\bk} = \frac{1}{2} \sqrt{\left(Jz+J_{\perp}\right)^{2}- \left(\alpha Jz\gamma_{\bk} \pm \alpha_{\perp}  J_{\perp}  \right)^{2}}.
\end{equation}

\section{Sum rule for $A_{\ro}$} \label{app:SumRule}
By following the same approach as presented in \cite{bruus2004} it is shown that 
\begin{align}
	\int_{-\infty}^{\infty}\frac{\text{d}\omega}{2\pi} {A}_{\rm o}(\bp,\omega) = 0.
	\label{Eq:Sum_rule}
\end{align}
From the definition of the off-diagonal spectral function
\begin{align}
	\int_{-\infty}^{\infty}\frac{\text{d}\omega}{2\pi} {A}_{\rm o}(\bp,\omega) &= -\int_{-\infty}^{\infty}\frac{\text{d}\omega}{2\pi}2\text{Im}[G^{R}_{\rm o}(\bp,\omega)]. 
	\label{Eq:Ao_sum_1}
\end{align}
As shown in \cite{bruus2004} for the diagonal part, one can by using the Lehmann representation state the imaginary part of the retarded Green's function as
\begin{align}
	2\text{Im}[G^{R}_{\rm o}(\bp,\omega)] =&- \frac{2\pi}{Z} \sum_{n,n'} \bra{n} \hat{h}_{2,\bp} \ket{n'}  \bra{n'} \hat{h}_{1,\bp}^{\dagger} \ket{n} \nonumber \\ 
	&\left(  e^{-\beta \rm{ E_{n}} } +  e^{-\beta \rm{ E_{n'}} }  \right) \delta(\omega + \rm{ E_{n}} - \rm{ E_{n'}}),
\end{align}
where $Z$ is a normalization constant defined such that $\left\langle ... \right\rangle = \sum_{n}  \bra{n} e^{-\beta \mathrm{\hat{H}}} ... \ket{n}/Z$, and the summation runs over all states with no holes present. Inserting this expression into Eq. \eqref{Eq:Ao_sum_1} we find
\begin{align}
	\int_{-\infty}^{\infty}\frac{\text{d}\omega}{2\pi} {A}_{\rm o}(\bp,\omega) =& \ \frac{1}{Z} \sum_{n,n'} \bra{n} \hat{h}_{\bp,2} \ket{n'}  \bra{n'} \hat{h}_{\bp,1}^{\dagger} \ket{n} \nonumber \\ 
	& \ \left(  e^{-\beta \rm{ E_{n}} } +  e^{-\beta \rm{ E_{n'}} }  \right) \nonumber \\
	=& \ \frac{1}{Z}\sum_{n,n'} \Big( \bra{n} \hat{h}_{2,\bp} \ket{n'}  \bra{n'} \hat{h}_{1,\bp}^{\dagger} \ket{n} e^{-\beta \rm{ E_{n} } }  \nonumber  \\  
	& \ + \bra{n'} \hat{h}_{1,\bp}^{\dagger} \ket{n}  \bra{n}  \hat{h}_{2,\bp}  \ket{n'} e^{-\beta \rm{ E_{n'}} } \Big) \nonumber  \\  
	=& \ \frac{1}{Z}\sum_{n} \Big( \bra{n} \hat{h}_{2,\bp} \hat{h}_{1,\bp}^{\dagger} \ket{n} e^{-\beta \rm{ E_{n} } }  \nonumber  \\  
	& \  + \bra{n} \hat{h}_{1,\bp}^{\dagger} \hat{h}_{2,\bp}  \ket{n} e^{-\beta \rm{ E_{n}} } \Big)   \nonumber  \\ 
	=& \ \left\langle  \hat{h}_{2,\bp} \hat{h}_{1,\bp}^{\dagger} + \hat{h}_{1,\bp}^{\dagger} \hat{h}_{2,\bp}   \right\rangle   \nonumber  \\ 
	=& \ 0. 
\end{align}

\bibliography{ArticleBilayer}

\begin{thebibliography}{49}%
\makeatletter
\providecommand \@ifxundefined [1]{%
 \@ifx{#1\undefined}
}%
\providecommand \@ifnum [1]{%
 \ifnum #1\expandafter \@firstoftwo
 \else \expandafter \@secondoftwo
 \fi
}%
\providecommand \@ifx [1]{%
 \ifx #1\expandafter \@firstoftwo
 \else \expandafter \@secondoftwo
 \fi
}%
\providecommand \natexlab [1]{#1}%
\providecommand \enquote  [1]{``#1''}%
\providecommand \bibnamefont  [1]{#1}%
\providecommand \bibfnamefont [1]{#1}%
\providecommand \citenamefont [1]{#1}%
\providecommand \href@noop [0]{\@secondoftwo}%
\providecommand \href [0]{\begingroup \@sanitize@url \@href}%
\providecommand \@href[1]{\@@startlink{#1}\@@href}%
\providecommand \@@href[1]{\endgroup#1\@@endlink}%
\providecommand \@sanitize@url [0]{\catcode `\\12\catcode `\$12\catcode
  `\&12\catcode `\#12\catcode `\^12\catcode `\_12\catcode `\%12\relax}%
\providecommand \@@startlink[1]{}%
\providecommand \@@endlink[0]{}%
\providecommand \url  [0]{\begingroup\@sanitize@url \@url }%
\providecommand \@url [1]{\endgroup\@href {#1}{\urlprefix }}%
\providecommand \urlprefix  [0]{URL }%
\providecommand \Eprint [0]{\href }%
\providecommand \doibase [0]{http://dx.doi.org/}%
\providecommand \selectlanguage [0]{\@gobble}%
\providecommand \bibinfo  [0]{\@secondoftwo}%
\providecommand \bibfield  [0]{\@secondoftwo}%
\providecommand \translation [1]{[#1]}%
\providecommand \BibitemOpen [0]{}%
\providecommand \bibitemStop [0]{}%
\providecommand \bibitemNoStop [0]{.\EOS\space}%
\providecommand \EOS [0]{\spacefactor3000\relax}%
\providecommand \BibitemShut  [1]{\csname bibitem#1\endcsname}%
\let\auto@bib@innerbib\@empty
\bibitem [{\citenamefont {Keimer}\ \emph {et~al.}(2015)\citenamefont {Keimer},
  \citenamefont {Kivelson}, \citenamefont {Norman}, \citenamefont {Uchida},\
  and\ \citenamefont {Zaanen}}]{Keimer:2015wh}%
  \BibitemOpen
  \bibfield  {author} {\bibinfo {author} {\bibfnamefont {B.}~\bibnamefont
  {Keimer}}, \bibinfo {author} {\bibfnamefont {S.~A.}\ \bibnamefont
  {Kivelson}}, \bibinfo {author} {\bibfnamefont {M.~R.}\ \bibnamefont
  {Norman}}, \bibinfo {author} {\bibfnamefont {S.}~\bibnamefont {Uchida}}, \
  and\ \bibinfo {author} {\bibfnamefont {J.}~\bibnamefont {Zaanen}},\
  }\bibfield  {title} {\enquote {\bibinfo {title} {From quantum matter to
  high-temperature superconductivity in copper oxides},}\ }\href {\doibase
  10.1038/nature14165} {\bibfield  {journal} {\bibinfo  {journal} {Nature}\
  }\textbf {\bibinfo {volume} {518}},\ \bibinfo {pages} {179--186} (\bibinfo
  {year} {2015})}\BibitemShut {NoStop}%
\bibitem [{\citenamefont {Wen}\ and\ \citenamefont {Li}(2011)}]{Wen2011}%
  \BibitemOpen
  \bibfield  {author} {\bibinfo {author} {\bibfnamefont {Hai-Hu}\ \bibnamefont
  {Wen}}\ and\ \bibinfo {author} {\bibfnamefont {Shiliang}\ \bibnamefont
  {Li}},\ }\bibfield  {title} {\enquote {\bibinfo {title} {Materials and novel
  superconductivity in iron pnictide superconductors},}\ }\href {\doibase
  10.1146/annurev-conmatphys-062910-140518} {\bibfield  {journal} {\bibinfo
  {journal} {Annual Review of Condensed Matter Physics}\ }\textbf {\bibinfo
  {volume} {2}},\ \bibinfo {pages} {121--140} (\bibinfo {year} {2011})},\
  \Eprint
  {http://arxiv.org/abs/https://doi.org/10.1146/annurev-conmatphys-062910-140518}
  {https://doi.org/10.1146/annurev-conmatphys-062910-140518} \BibitemShut
  {NoStop}%
\bibitem [{\citenamefont {Wosnitza}(2012)}]{Wosnitza2012}%
  \BibitemOpen
  \bibfield  {author} {\bibinfo {author} {\bibfnamefont {Jochen}\ \bibnamefont
  {Wosnitza}},\ }\bibfield  {title} {\enquote {\bibinfo {title}
  {Superconductivity in layered organic metals},}\ }\href {\doibase
  10.3390/cryst2020248} {\bibfield  {journal} {\bibinfo  {journal} {Crystals}\
  }\textbf {\bibinfo {volume} {2}},\ \bibinfo {pages} {248--265} (\bibinfo
  {year} {2012})}\BibitemShut {NoStop}%
\bibitem [{\citenamefont {Cao}\ \emph {et~al.}(2018)\citenamefont {Cao},
  \citenamefont {Fatemi}, \citenamefont {Fang}, \citenamefont {Watanabe},
  \citenamefont {Taniguchi}, \citenamefont {Kaxiras},\ and\ \citenamefont
  {Jarillo-Herrero}}]{Cao:2018wy}%
  \BibitemOpen
  \bibfield  {author} {\bibinfo {author} {\bibfnamefont {Yuan}\ \bibnamefont
  {Cao}}, \bibinfo {author} {\bibfnamefont {Valla}\ \bibnamefont {Fatemi}},
  \bibinfo {author} {\bibfnamefont {Shiang}\ \bibnamefont {Fang}}, \bibinfo
  {author} {\bibfnamefont {Kenji}\ \bibnamefont {Watanabe}}, \bibinfo {author}
  {\bibfnamefont {Takashi}\ \bibnamefont {Taniguchi}}, \bibinfo {author}
  {\bibfnamefont {Efthimios}\ \bibnamefont {Kaxiras}}, \ and\ \bibinfo {author}
  {\bibfnamefont {Pablo}\ \bibnamefont {Jarillo-Herrero}},\ }\bibfield  {title}
  {\enquote {\bibinfo {title} {Unconventional superconductivity in magic-angle
  graphene superlattices},}\ }\href {\doibase 10.1038/nature26160} {\bibfield
  {journal} {\bibinfo  {journal} {Nature}\ }\textbf {\bibinfo {volume} {556}},\
  \bibinfo {pages} {43--50} (\bibinfo {year} {2018})}\BibitemShut {NoStop}%
\bibitem [{\citenamefont {Dagotto}(1994)}]{Dagotto1994}%
  \BibitemOpen
  \bibfield  {author} {\bibinfo {author} {\bibfnamefont {Elbio}\ \bibnamefont
  {Dagotto}},\ }\bibfield  {title} {\enquote {\bibinfo {title} {Correlated
  electrons in high-temperature superconductors},}\ }\href {\doibase
  10.1103/RevModPhys.66.763} {\bibfield  {journal} {\bibinfo  {journal} {Rev.
  Mod. Phys.}\ }\textbf {\bibinfo {volume} {66}},\ \bibinfo {pages} {763--840}
  (\bibinfo {year} {1994})}\BibitemShut {NoStop}%
\bibitem [{\citenamefont {Lee}\ \emph {et~al.}(2006)\citenamefont {Lee},
  \citenamefont {Nagaosa},\ and\ \citenamefont {Wen}}]{Lee2006}%
  \BibitemOpen
  \bibfield  {author} {\bibinfo {author} {\bibfnamefont {Patrick~A.}\
  \bibnamefont {Lee}}, \bibinfo {author} {\bibfnamefont {Naoto}\ \bibnamefont
  {Nagaosa}}, \ and\ \bibinfo {author} {\bibfnamefont {Xiao-Gang}\ \bibnamefont
  {Wen}},\ }\bibfield  {title} {\enquote {\bibinfo {title} {Doping a mott
  insulator: Physics of high-temperature superconductivity},}\ }\href {\doibase
  10.1103/RevModPhys.78.17} {\bibfield  {journal} {\bibinfo  {journal} {Rev.
  Mod. Phys.}\ }\textbf {\bibinfo {volume} {78}},\ \bibinfo {pages} {17--85}
  (\bibinfo {year} {2006})}\BibitemShut {NoStop}%
\bibitem [{\citenamefont {Damascelli}\ \emph {et~al.}(2003)\citenamefont
  {Damascelli}, \citenamefont {Hussain},\ and\ \citenamefont
  {Shen}}]{Damascelli2003}%
  \BibitemOpen
  \bibfield  {author} {\bibinfo {author} {\bibfnamefont {Andrea}\ \bibnamefont
  {Damascelli}}, \bibinfo {author} {\bibfnamefont {Zahid}\ \bibnamefont
  {Hussain}}, \ and\ \bibinfo {author} {\bibfnamefont {Zhi-Xun}\ \bibnamefont
  {Shen}},\ }\bibfield  {title} {\enquote {\bibinfo {title} {Angle-resolved
  photoemission studies of the cuprate superconductors},}\ }\href {\doibase
  10.1103/RevModPhys.75.473} {\bibfield  {journal} {\bibinfo  {journal} {Rev.
  Mod. Phys.}\ }\textbf {\bibinfo {volume} {75}},\ \bibinfo {pages} {473--541}
  (\bibinfo {year} {2003})}\BibitemShut {NoStop}%
\bibitem [{\citenamefont {{Schmitt-Rink}}\ \emph {et~al.}(1988)\citenamefont
  {{Schmitt-Rink}}, \citenamefont {Varma},\ and\ \citenamefont
  {Ruckenstein}}]{schmitt-rink1988}%
  \BibitemOpen
  \bibfield  {author} {\bibinfo {author} {\bibfnamefont {S.}~\bibnamefont
  {{Schmitt-Rink}}}, \bibinfo {author} {\bibfnamefont {C.~M.}\ \bibnamefont
  {Varma}}, \ and\ \bibinfo {author} {\bibfnamefont {A.~E.}\ \bibnamefont
  {Ruckenstein}},\ }\bibfield  {title} {\enquote {\bibinfo {title} {Spectral
  {{Function}} of {{Holes}} in a {{Quantum Antiferromagnet}}},}\ }\href
  {\doibase 10.1103/PhysRevLett.60.2793} {\bibfield  {journal} {\bibinfo
  {journal} {Physical Review Letters}\ }\textbf {\bibinfo {volume} {60}},\
  \bibinfo {pages} {2793--2796} (\bibinfo {year} {1988})}\BibitemShut {NoStop}%
\bibitem [{\citenamefont {Shraiman}\ and\ \citenamefont
  {Siggia}(1988)}]{shraiman1988}%
  \BibitemOpen
  \bibfield  {author} {\bibinfo {author} {\bibfnamefont {Boris~I.}\
  \bibnamefont {Shraiman}}\ and\ \bibinfo {author} {\bibfnamefont {Eric~D.}\
  \bibnamefont {Siggia}},\ }\bibfield  {title} {\enquote {\bibinfo {title}
  {Mobile {{Vacancies}} in a {{Quantum Heisenberg Antiferromagnet}}},}\ }\href
  {\doibase 10.1103/PhysRevLett.61.467} {\bibfield  {journal} {\bibinfo
  {journal} {Physical Review Letters}\ }\textbf {\bibinfo {volume} {61}},\
  \bibinfo {pages} {467--470} (\bibinfo {year} {1988})}\BibitemShut {NoStop}%
\bibitem [{\citenamefont {Kane}\ \emph {et~al.}(1989)\citenamefont {Kane},
  \citenamefont {Lee},\ and\ \citenamefont {Read}}]{kane1989}%
  \BibitemOpen
  \bibfield  {author} {\bibinfo {author} {\bibfnamefont {C.~L.}\ \bibnamefont
  {Kane}}, \bibinfo {author} {\bibfnamefont {P.~A.}\ \bibnamefont {Lee}}, \
  and\ \bibinfo {author} {\bibfnamefont {N.}~\bibnamefont {Read}},\ }\bibfield
  {title} {\enquote {\bibinfo {title} {Motion of a single hole in a quantum
  antiferromagnet},}\ }\href {\doibase 10.1103/PhysRevB.39.6880} {\bibfield
  {journal} {\bibinfo  {journal} {Physical Review B}\ }\textbf {\bibinfo
  {volume} {39}},\ \bibinfo {pages} {6880--6897} (\bibinfo {year}
  {1989})}\BibitemShut {NoStop}%
\bibitem [{\citenamefont {Martinez}\ and\ \citenamefont
  {Horsch}(1991)}]{martinez1991}%
  \BibitemOpen
  \bibfield  {author} {\bibinfo {author} {\bibfnamefont {Gerardo}\ \bibnamefont
  {Martinez}}\ and\ \bibinfo {author} {\bibfnamefont {Peter}\ \bibnamefont
  {Horsch}},\ }\bibfield  {title} {\enquote {\bibinfo {title} {Spin polarons in
  the {\emph{t}} - {{{\emph{J}}}} model},}\ }\href {\doibase
  10.1103/PhysRevB.44.317} {\bibfield  {journal} {\bibinfo  {journal} {Physical
  Review B}\ }\textbf {\bibinfo {volume} {44}},\ \bibinfo {pages} {317--331}
  (\bibinfo {year} {1991})}\BibitemShut {NoStop}%
\bibitem [{\citenamefont {Liu}\ and\ \citenamefont
  {Manousakis}(1991)}]{liu1991}%
  \BibitemOpen
  \bibfield  {author} {\bibinfo {author} {\bibfnamefont {Zhiping}\ \bibnamefont
  {Liu}}\ and\ \bibinfo {author} {\bibfnamefont {Efstratios}\ \bibnamefont
  {Manousakis}},\ }\bibfield  {title} {\enquote {\bibinfo {title} {Spectral
  function of a hole in the {\emph{t}} - {{{\emph{J}}}} model},}\ }\href
  {\doibase 10.1103/PhysRevB.44.2414} {\bibfield  {journal} {\bibinfo
  {journal} {Physical Review B}\ }\textbf {\bibinfo {volume} {44}},\ \bibinfo
  {pages} {2414--2417} (\bibinfo {year} {1991})}\BibitemShut {NoStop}%
\bibitem [{\citenamefont {Marsiglio}\ \emph {et~al.}(1991)\citenamefont
  {Marsiglio}, \citenamefont {Ruckenstein}, \citenamefont {{Schmitt-Rink}},\
  and\ \citenamefont {Varma}}]{marsiglio1991}%
  \BibitemOpen
  \bibfield  {author} {\bibinfo {author} {\bibfnamefont {Frank}\ \bibnamefont
  {Marsiglio}}, \bibinfo {author} {\bibfnamefont {Andrei~E.}\ \bibnamefont
  {Ruckenstein}}, \bibinfo {author} {\bibfnamefont {Stefan}\ \bibnamefont
  {{Schmitt-Rink}}}, \ and\ \bibinfo {author} {\bibfnamefont {Chandra~M.}\
  \bibnamefont {Varma}},\ }\bibfield  {title} {\enquote {\bibinfo {title}
  {Spectral function of a single hole in a two-dimensional quantum
  antiferromagnet},}\ }\href {\doibase 10.1103/PhysRevB.43.10882} {\bibfield
  {journal} {\bibinfo  {journal} {Physical Review B}\ }\textbf {\bibinfo
  {volume} {43}},\ \bibinfo {pages} {10882--10889} (\bibinfo {year}
  {1991})}\BibitemShut {NoStop}%
\bibitem [{\citenamefont {Chernyshev}\ and\ \citenamefont
  {Leung}(1999)}]{chernyshev1999}%
  \BibitemOpen
  \bibfield  {author} {\bibinfo {author} {\bibfnamefont {A.~L.}\ \bibnamefont
  {Chernyshev}}\ and\ \bibinfo {author} {\bibfnamefont {P.~W.}\ \bibnamefont
  {Leung}},\ }\bibfield  {title} {\enquote {\bibinfo {title} {Holes in the t -
  {{J}} z model: {{A}} diagrammatic study},}\ }\href {\doibase
  10.1103/PhysRevB.60.1592} {\bibfield  {journal} {\bibinfo  {journal}
  {Physical Review B}\ }\textbf {\bibinfo {volume} {60}},\ \bibinfo {pages}
  {1592--1606} (\bibinfo {year} {1999})}\BibitemShut {NoStop}%
\bibitem [{\citenamefont {Diamantis}\ and\ \citenamefont
  {Manousakis}(2021)}]{Diamantis_2021}%
  \BibitemOpen
  \bibfield  {author} {\bibinfo {author} {\bibfnamefont {Nikolaos~G}\
  \bibnamefont {Diamantis}}\ and\ \bibinfo {author} {\bibfnamefont
  {Efstratios}\ \bibnamefont {Manousakis}},\ }\bibfield  {title} {\enquote
  {\bibinfo {title} {Dynamics of string-like states of a hole in a quantum
  antiferromagnet: a diagrammatic monte carlo simulation},}\ }\href {\doibase
  10.1088/1367-2630/ac39b5} {\bibfield  {journal} {\bibinfo  {journal} {New
  Journal of Physics}\ }\textbf {\bibinfo {volume} {23}},\ \bibinfo {pages}
  {123005} (\bibinfo {year} {2021})}\BibitemShut {NoStop}%
\bibitem [{\citenamefont {Nielsen}\ \emph {et~al.}(2022)\citenamefont
  {Nielsen}, \citenamefont {Pohl},\ and\ \citenamefont {Bruun}}]{nielsen2022}%
  \BibitemOpen
  \bibfield  {author} {\bibinfo {author} {\bibfnamefont {K.~Knakkergaard}\
  \bibnamefont {Nielsen}}, \bibinfo {author} {\bibfnamefont {T.}~\bibnamefont
  {Pohl}}, \ and\ \bibinfo {author} {\bibfnamefont {G.~M.}\ \bibnamefont
  {Bruun}},\ }\bibfield  {title} {\enquote {\bibinfo {title} {Non-equilibrium
  hole dynamics in antiferromagnets: Damped strings and polarons},}\ }\href
  {\doibase 10.48550/arXiv.2203.04789} {\bibfield  {journal} {\bibinfo
  {journal} {arXiv:2203.04789v1 [cond-mat.quant-gas]}\ } (\bibinfo {year}
  {2022}),\ 10.48550/arXiv.2203.04789},\ \Eprint
  {http://arxiv.org/abs/2203.04789v1} {arXiv:2203.04789v1 [cond-mat.quant-gas]}
  \BibitemShut {NoStop}%
\bibitem [{\citenamefont {Nazarenko}\ and\ \citenamefont
  {Dagotto}(1996)}]{nazarenko1996}%
  \BibitemOpen
  \bibfield  {author} {\bibinfo {author} {\bibfnamefont {Alexander}\
  \bibnamefont {Nazarenko}}\ and\ \bibinfo {author} {\bibfnamefont {Elbio}\
  \bibnamefont {Dagotto}},\ }\bibfield  {title} {\enquote {\bibinfo {title}
  {Hole dispersion and symmetry of the superconducting order parameter for
  underdoped {{CuO}} 2 bilayers and the three-dimensional antiferromagnets},}\
  }\href {\doibase 10.1103/PhysRevB.54.13158} {\bibfield  {journal} {\bibinfo
  {journal} {Physical Review B}\ }\textbf {\bibinfo {volume} {54}},\ \bibinfo
  {pages} {13158--13166} (\bibinfo {year} {1996})}\BibitemShut {NoStop}%
\bibitem [{\citenamefont {Yin}\ and\ \citenamefont {Gong}(1997)}]{yin1997}%
  \BibitemOpen
  \bibfield  {author} {\bibinfo {author} {\bibfnamefont {Wei-Guo}\ \bibnamefont
  {Yin}}\ and\ \bibinfo {author} {\bibfnamefont {Chang-De}\ \bibnamefont
  {Gong}},\ }\bibfield  {title} {\enquote {\bibinfo {title} {Quasiparticle
  bands in the realistic bilayer cuprates},}\ }\href {\doibase
  10.1103/PhysRevB.56.2843} {\bibfield  {journal} {\bibinfo  {journal}
  {Physical Review B}\ }\textbf {\bibinfo {volume} {56}},\ \bibinfo {pages}
  {2843--2846} (\bibinfo {year} {1997})}\BibitemShut {NoStop}%
\bibitem [{\citenamefont {Yin}\ and\ \citenamefont {Gong}(1998)}]{yin1998}%
  \BibitemOpen
  \bibfield  {author} {\bibinfo {author} {\bibfnamefont {Wei-Guo}\ \bibnamefont
  {Yin}}\ and\ \bibinfo {author} {\bibfnamefont {Chang-De}\ \bibnamefont
  {Gong}},\ }\bibfield  {title} {\enquote {\bibinfo {title} {Quasiparticle
  bands and superconductivity for the multiple-layer and three-dimensional
  superlattice t - {{J}} models},}\ }\href {\doibase 10.1103/PhysRevB.57.11743}
  {\bibfield  {journal} {\bibinfo  {journal} {Physical Review B}\ }\textbf
  {\bibinfo {volume} {57}},\ \bibinfo {pages} {11743--11751} (\bibinfo {year}
  {1998})}\BibitemShut {NoStop}%
\bibitem [{\citenamefont {Hida}(1992)}]{Hida1992}%
  \BibitemOpen
  \bibfield  {author} {\bibinfo {author} {\bibfnamefont {Kazuo}\ \bibnamefont
  {Hida}},\ }\bibfield  {title} {\enquote {\bibinfo {title} {Quantum disordered
  state without frustration in the double layer heisenberg antiferromagnet
  --dimer expansion and projector monte carlo study--},}\ }\href {\doibase
  10.1143/JPSJ.61.1013} {\bibfield  {journal} {\bibinfo  {journal} {Journal of
  the Physical Society of Japan}\ }\textbf {\bibinfo {volume} {61}},\ \bibinfo
  {pages} {1013--1018} (\bibinfo {year} {1992})},\ \Eprint
  {http://arxiv.org/abs/https://doi.org/10.1143/JPSJ.61.1013}
  {https://doi.org/10.1143/JPSJ.61.1013} \BibitemShut {NoStop}%
\bibitem [{\citenamefont {Sandvik}\ and\ \citenamefont
  {Scalapino}(1994)}]{Sandvik1994}%
  \BibitemOpen
  \bibfield  {author} {\bibinfo {author} {\bibfnamefont {A.~W.}\ \bibnamefont
  {Sandvik}}\ and\ \bibinfo {author} {\bibfnamefont {D.~J.}\ \bibnamefont
  {Scalapino}},\ }\bibfield  {title} {\enquote {\bibinfo {title}
  {Order-disorder transition in a two-layer quantum antiferromagnet},}\ }\href
  {\doibase 10.1103/PhysRevLett.72.2777} {\bibfield  {journal} {\bibinfo
  {journal} {Phys. Rev. Lett.}\ }\textbf {\bibinfo {volume} {72}},\ \bibinfo
  {pages} {2777--2780} (\bibinfo {year} {1994})}\BibitemShut {NoStop}%
\bibitem [{\citenamefont {Scalettar}\ \emph {et~al.}(1994)\citenamefont
  {Scalettar}, \citenamefont {Cannon}, \citenamefont {Scalapino},\ and\
  \citenamefont {Sugar}}]{scalettar1994}%
  \BibitemOpen
  \bibfield  {author} {\bibinfo {author} {\bibfnamefont {Richard~T.}\
  \bibnamefont {Scalettar}}, \bibinfo {author} {\bibfnamefont {Joel~W.}\
  \bibnamefont {Cannon}}, \bibinfo {author} {\bibfnamefont {Douglas~J.}\
  \bibnamefont {Scalapino}}, \ and\ \bibinfo {author} {\bibfnamefont
  {Robert~L.}\ \bibnamefont {Sugar}},\ }\bibfield  {title} {\enquote {\bibinfo
  {title} {Magnetic and pairing correlations in coupled {{Hubbard}} planes},}\
  }\href {\doibase 10.1103/PhysRevB.50.13419} {\bibfield  {journal} {\bibinfo
  {journal} {Physical Review B}\ }\textbf {\bibinfo {volume} {50}},\ \bibinfo
  {pages} {13419--13427} (\bibinfo {year} {1994})}\BibitemShut {NoStop}%
\bibitem [{\citenamefont {Millis}\ and\ \citenamefont
  {Monien}(1994)}]{millis1994}%
  \BibitemOpen
  \bibfield  {author} {\bibinfo {author} {\bibfnamefont {A.~J.}\ \bibnamefont
  {Millis}}\ and\ \bibinfo {author} {\bibfnamefont {H.}~\bibnamefont
  {Monien}},\ }\bibfield  {title} {\enquote {\bibinfo {title} {Spin gaps and
  bilayer coupling in {{YBa}} 2 {{Cu}} 3 {{O}} 7 - {$\delta$} and {{YBa}} 2
  {{Cu}} 4 {{O}} 8},}\ }\href {\doibase 10.1103/PhysRevB.50.16606} {\bibfield
  {journal} {\bibinfo  {journal} {Physical Review B}\ }\textbf {\bibinfo
  {volume} {50}},\ \bibinfo {pages} {16606--16622} (\bibinfo {year}
  {1994})}\BibitemShut {NoStop}%
\bibitem [{\citenamefont {Sandvik}\ \emph {et~al.}(1995)\citenamefont
  {Sandvik}, \citenamefont {Chubukov},\ and\ \citenamefont
  {Sachdev}}]{Sandvik1995}%
  \BibitemOpen
  \bibfield  {author} {\bibinfo {author} {\bibfnamefont {Anders~W.}\
  \bibnamefont {Sandvik}}, \bibinfo {author} {\bibfnamefont {Andrey~V.}\
  \bibnamefont {Chubukov}}, \ and\ \bibinfo {author} {\bibfnamefont {Subir}\
  \bibnamefont {Sachdev}},\ }\bibfield  {title} {\enquote {\bibinfo {title}
  {Quantum critical behavior in a two-layer antiferromagnet},}\ }\href
  {\doibase 10.1103/PhysRevB.51.16483} {\bibfield  {journal} {\bibinfo
  {journal} {Phys. Rev. B}\ }\textbf {\bibinfo {volume} {51}},\ \bibinfo
  {pages} {16483--16486} (\bibinfo {year} {1995})}\BibitemShut {NoStop}%
\bibitem [{\citenamefont {Chubukov}\ and\ \citenamefont
  {Morr}(1995)}]{chubukov1995}%
  \BibitemOpen
  \bibfield  {author} {\bibinfo {author} {\bibfnamefont {Andrey~V.}\
  \bibnamefont {Chubukov}}\ and\ \bibinfo {author} {\bibfnamefont {Dirk~K.}\
  \bibnamefont {Morr}},\ }\bibfield  {title} {\enquote {\bibinfo {title} {Phase
  transition, longitudinal spin fluctuations, and scaling in a two-layer
  antiferromagnet},}\ }\href {\doibase 10.1103/PhysRevB.52.3521} {\bibfield
  {journal} {\bibinfo  {journal} {Physical Review B}\ }\textbf {\bibinfo
  {volume} {52}},\ \bibinfo {pages} {3521--3532} (\bibinfo {year}
  {1995})}\BibitemShut {NoStop}%
\bibitem [{\citenamefont {Carlstr\"om}\ \emph {et~al.}(2016)\citenamefont
  {Carlstr\"om}, \citenamefont {Prokof'ev},\ and\ \citenamefont
  {Svistunov}}]{Carlstrom2016}%
  \BibitemOpen
  \bibfield  {author} {\bibinfo {author} {\bibfnamefont {Johan}\ \bibnamefont
  {Carlstr\"om}}, \bibinfo {author} {\bibfnamefont {Nikolay}\ \bibnamefont
  {Prokof'ev}}, \ and\ \bibinfo {author} {\bibfnamefont {Boris}\ \bibnamefont
  {Svistunov}},\ }\bibfield  {title} {\enquote {\bibinfo {title} {Quantum walk
  in degenerate spin environments},}\ }\href {\doibase
  10.1103/PhysRevLett.116.247202} {\bibfield  {journal} {\bibinfo  {journal}
  {Phys. Rev. Lett.}\ }\textbf {\bibinfo {volume} {116}},\ \bibinfo {pages}
  {247202} (\bibinfo {year} {2016})}\BibitemShut {NoStop}%
\bibitem [{\citenamefont {Kan\'asz-Nagy}\ \emph {et~al.}(2017)\citenamefont
  {Kan\'asz-Nagy}, \citenamefont {Lovas}, \citenamefont {Grusdt}, \citenamefont
  {Greif}, \citenamefont {Greiner},\ and\ \citenamefont {Demler}}]{Nagy2017}%
  \BibitemOpen
  \bibfield  {author} {\bibinfo {author} {\bibfnamefont {M\'arton}\
  \bibnamefont {Kan\'asz-Nagy}}, \bibinfo {author} {\bibfnamefont {Izabella}\
  \bibnamefont {Lovas}}, \bibinfo {author} {\bibfnamefont {Fabian}\
  \bibnamefont {Grusdt}}, \bibinfo {author} {\bibfnamefont {Daniel}\
  \bibnamefont {Greif}}, \bibinfo {author} {\bibfnamefont {Markus}\
  \bibnamefont {Greiner}}, \ and\ \bibinfo {author} {\bibfnamefont {Eugene~A.}\
  \bibnamefont {Demler}},\ }\bibfield  {title} {\enquote {\bibinfo {title}
  {Quantum correlations at infinite temperature: The dynamical nagaoka
  effect},}\ }\href {\doibase 10.1103/PhysRevB.96.014303} {\bibfield  {journal}
  {\bibinfo  {journal} {Phys. Rev. B}\ }\textbf {\bibinfo {volume} {96}},\
  \bibinfo {pages} {014303} (\bibinfo {year} {2017})}\BibitemShut {NoStop}%
\bibitem [{\citenamefont {Grusdt}\ \emph
  {et~al.}(2018{\natexlab{a}})\citenamefont {Grusdt}, \citenamefont {Zhu},
  \citenamefont {Shi},\ and\ \citenamefont {Demler}}]{grusdt2018}%
  \BibitemOpen
  \bibfield  {author} {\bibinfo {author} {\bibfnamefont {Fabian}\ \bibnamefont
  {Grusdt}}, \bibinfo {author} {\bibfnamefont {Zheng}\ \bibnamefont {Zhu}},
  \bibinfo {author} {\bibfnamefont {Tao}\ \bibnamefont {Shi}}, \ and\ \bibinfo
  {author} {\bibfnamefont {Eugene}\ \bibnamefont {Demler}},\ }\bibfield
  {title} {\enquote {\bibinfo {title} {Meson formation in mixed-dimensional
  t-{{J}} models},}\ }\href {\doibase 10.21468/SciPostPhys.5.6.057} {\bibfield
  {journal} {\bibinfo  {journal} {SciPost Physics}\ }\textbf {\bibinfo {volume}
  {5}},\ \bibinfo {pages} {057} (\bibinfo {year}
  {2018}{\natexlab{a}})}\BibitemShut {NoStop}%
\bibitem [{\citenamefont {Grusdt}\ \emph
  {et~al.}(2018{\natexlab{b}})\citenamefont {Grusdt}, \citenamefont
  {K\'anasz-Nagy}, \citenamefont {Bohrdt}, \citenamefont {Chiu}, \citenamefont
  {Ji}, \citenamefont {Greiner}, \citenamefont {Greif},\ and\ \citenamefont
  {Demler}}]{Grusdt2018b}%
  \BibitemOpen
  \bibfield  {author} {\bibinfo {author} {\bibfnamefont {F.}~\bibnamefont
  {Grusdt}}, \bibinfo {author} {\bibfnamefont {M.}~\bibnamefont
  {K\'anasz-Nagy}}, \bibinfo {author} {\bibfnamefont {A.}~\bibnamefont
  {Bohrdt}}, \bibinfo {author} {\bibfnamefont {C.~S.}\ \bibnamefont {Chiu}},
  \bibinfo {author} {\bibfnamefont {G.}~\bibnamefont {Ji}}, \bibinfo {author}
  {\bibfnamefont {M.}~\bibnamefont {Greiner}}, \bibinfo {author} {\bibfnamefont
  {D.}~\bibnamefont {Greif}}, \ and\ \bibinfo {author} {\bibfnamefont
  {E.}~\bibnamefont {Demler}},\ }\bibfield  {title} {\enquote {\bibinfo {title}
  {Parton theory of magnetic polarons: Mesonic resonances and signatures in
  dynamics},}\ }\href {\doibase 10.1103/PhysRevX.8.011046} {\bibfield
  {journal} {\bibinfo  {journal} {Phys. Rev. X}\ }\textbf {\bibinfo {volume}
  {8}},\ \bibinfo {pages} {011046} (\bibinfo {year}
  {2018}{\natexlab{b}})}\BibitemShut {NoStop}%
\bibitem [{\citenamefont {Nielsen}\ \emph {et~al.}(2021)\citenamefont
  {Nielsen}, \citenamefont {{Bastarrachea-Magnani}}, \citenamefont {Pohl},\
  and\ \citenamefont {Bruun}}]{nielsen2021}%
  \BibitemOpen
  \bibfield  {author} {\bibinfo {author} {\bibfnamefont {K.~K.}\ \bibnamefont
  {Nielsen}}, \bibinfo {author} {\bibfnamefont {M.~A.}\ \bibnamefont
  {{Bastarrachea-Magnani}}}, \bibinfo {author} {\bibfnamefont {T.}~\bibnamefont
  {Pohl}}, \ and\ \bibinfo {author} {\bibfnamefont {G.~M.}\ \bibnamefont
  {Bruun}},\ }\bibfield  {title} {\enquote {\bibinfo {title} {Spatial structure
  of magnetic polarons in strongly interacting antiferromagnets},}\ }\href
  {\doibase 10.1103/PhysRevB.104.155136} {\bibfield  {journal} {\bibinfo
  {journal} {Physical Review B}\ }\textbf {\bibinfo {volume} {104}},\ \bibinfo
  {pages} {155136} (\bibinfo {year} {2021})}\BibitemShut {NoStop}%
\bibitem [{\citenamefont {Chiu}\ \emph {et~al.}(2019)\citenamefont {Chiu},
  \citenamefont {Ji}, \citenamefont {Bohrdt}, \citenamefont {Xu}, \citenamefont
  {Knap}, \citenamefont {Demler}, \citenamefont {Grusdt}, \citenamefont
  {Greiner},\ and\ \citenamefont {Greif}}]{Christie2019}%
  \BibitemOpen
  \bibfield  {author} {\bibinfo {author} {\bibfnamefont {Christie~S.}\
  \bibnamefont {Chiu}}, \bibinfo {author} {\bibfnamefont {Geoffrey}\
  \bibnamefont {Ji}}, \bibinfo {author} {\bibfnamefont {Annabelle}\
  \bibnamefont {Bohrdt}}, \bibinfo {author} {\bibfnamefont {Muqing}\
  \bibnamefont {Xu}}, \bibinfo {author} {\bibfnamefont {Michael}\ \bibnamefont
  {Knap}}, \bibinfo {author} {\bibfnamefont {Eugene}\ \bibnamefont {Demler}},
  \bibinfo {author} {\bibfnamefont {Fabian}\ \bibnamefont {Grusdt}}, \bibinfo
  {author} {\bibfnamefont {Markus}\ \bibnamefont {Greiner}}, \ and\ \bibinfo
  {author} {\bibfnamefont {Daniel}\ \bibnamefont {Greif}},\ }\bibfield  {title}
  {\enquote {\bibinfo {title} {String patterns in the doped hubbard model},}\
  }\href {\doibase 10.1126/science.aav3587} {\bibfield  {journal} {\bibinfo
  {journal} {Science}\ }\textbf {\bibinfo {volume} {365}},\ \bibinfo {pages}
  {251--256} (\bibinfo {year} {2019})},\ \Eprint
  {http://arxiv.org/abs/https://www.science.org/doi/pdf/10.1126/science.aav3587}
  {https://www.science.org/doi/pdf/10.1126/science.aav3587} \BibitemShut
  {NoStop}%
\bibitem [{\citenamefont {Brown}\ \emph {et~al.}(2019)\citenamefont {Brown},
  \citenamefont {Mitra}, \citenamefont {Guardado-Sanchez}, \citenamefont
  {Nourafkan}, \citenamefont {Reymbaut}, \citenamefont {H{\'e}bert},
  \citenamefont {Bergeron}, \citenamefont {Tremblay}, \citenamefont {Kokalj},
  \citenamefont {Huse}, \citenamefont {Schau{\ss}},\ and\ \citenamefont
  {Bakr}}]{Brown2019}%
  \BibitemOpen
  \bibfield  {author} {\bibinfo {author} {\bibfnamefont {Peter~T.}\
  \bibnamefont {Brown}}, \bibinfo {author} {\bibfnamefont {Debayan}\
  \bibnamefont {Mitra}}, \bibinfo {author} {\bibfnamefont {Elmer}\ \bibnamefont
  {Guardado-Sanchez}}, \bibinfo {author} {\bibfnamefont {Reza}\ \bibnamefont
  {Nourafkan}}, \bibinfo {author} {\bibfnamefont {Alexis}\ \bibnamefont
  {Reymbaut}}, \bibinfo {author} {\bibfnamefont {Charles-David}\ \bibnamefont
  {H{\'e}bert}}, \bibinfo {author} {\bibfnamefont {Simon}\ \bibnamefont
  {Bergeron}}, \bibinfo {author} {\bibfnamefont {A.-M.~S.}\ \bibnamefont
  {Tremblay}}, \bibinfo {author} {\bibfnamefont {Jure}\ \bibnamefont {Kokalj}},
  \bibinfo {author} {\bibfnamefont {David~A.}\ \bibnamefont {Huse}}, \bibinfo
  {author} {\bibfnamefont {Peter}\ \bibnamefont {Schau{\ss}}}, \ and\ \bibinfo
  {author} {\bibfnamefont {Waseem~S.}\ \bibnamefont {Bakr}},\ }\bibfield
  {title} {\enquote {\bibinfo {title} {Bad metallic transport in a cold atom
  fermi-hubbard system},}\ }\href {\doibase 10.1126/science.aat4134} {\bibfield
   {journal} {\bibinfo  {journal} {Science}\ }\textbf {\bibinfo {volume}
  {363}},\ \bibinfo {pages} {379--382} (\bibinfo {year} {2019})},\ \Eprint
  {http://arxiv.org/abs/https://www.science.org/doi/pdf/10.1126/science.aat4134}
  {https://www.science.org/doi/pdf/10.1126/science.aat4134} \BibitemShut
  {NoStop}%
\bibitem [{\citenamefont {Koepsell}\ \emph {et~al.}(2019)\citenamefont
  {Koepsell}, \citenamefont {Vijayan}, \citenamefont {Sompet}, \citenamefont
  {Grusdt}, \citenamefont {Hilker}, \citenamefont {Demler}, \citenamefont
  {Salomon}, \citenamefont {Bloch},\ and\ \citenamefont
  {Gross}}]{Koepsell:2019ua}%
  \BibitemOpen
  \bibfield  {author} {\bibinfo {author} {\bibfnamefont {Joannis}\ \bibnamefont
  {Koepsell}}, \bibinfo {author} {\bibfnamefont {Jayadev}\ \bibnamefont
  {Vijayan}}, \bibinfo {author} {\bibfnamefont {Pimonpan}\ \bibnamefont
  {Sompet}}, \bibinfo {author} {\bibfnamefont {Fabian}\ \bibnamefont {Grusdt}},
  \bibinfo {author} {\bibfnamefont {Timon~A.}\ \bibnamefont {Hilker}}, \bibinfo
  {author} {\bibfnamefont {Eugene}\ \bibnamefont {Demler}}, \bibinfo {author}
  {\bibfnamefont {Guillaume}\ \bibnamefont {Salomon}}, \bibinfo {author}
  {\bibfnamefont {Immanuel}\ \bibnamefont {Bloch}}, \ and\ \bibinfo {author}
  {\bibfnamefont {Christian}\ \bibnamefont {Gross}},\ }\bibfield  {title}
  {\enquote {\bibinfo {title} {Imaging magnetic polarons in the doped
  fermi--hubbard model},}\ }\href {\doibase 10.1038/s41586-019-1463-1}
  {\bibfield  {journal} {\bibinfo  {journal} {Nature}\ }\textbf {\bibinfo
  {volume} {572}},\ \bibinfo {pages} {358--362} (\bibinfo {year}
  {2019})}\BibitemShut {NoStop}%
\bibitem [{\citenamefont {Ji}\ \emph {et~al.}(2021)\citenamefont {Ji},
  \citenamefont {Xu}, \citenamefont {Kendrick}, \citenamefont {Chiu},
  \citenamefont {Br\"uggenj\"urgen}, \citenamefont {Greif}, \citenamefont
  {Bohrdt}, \citenamefont {Grusdt}, \citenamefont {Demler}, \citenamefont
  {Lebrat},\ and\ \citenamefont {Greiner}}]{Ji2021}%
  \BibitemOpen
  \bibfield  {author} {\bibinfo {author} {\bibfnamefont {Geoffrey}\
  \bibnamefont {Ji}}, \bibinfo {author} {\bibfnamefont {Muqing}\ \bibnamefont
  {Xu}}, \bibinfo {author} {\bibfnamefont {Lev~Haldar}\ \bibnamefont
  {Kendrick}}, \bibinfo {author} {\bibfnamefont {Christie~S.}\ \bibnamefont
  {Chiu}}, \bibinfo {author} {\bibfnamefont {Justus~C.}\ \bibnamefont
  {Br\"uggenj\"urgen}}, \bibinfo {author} {\bibfnamefont {Daniel}\ \bibnamefont
  {Greif}}, \bibinfo {author} {\bibfnamefont {Annabelle}\ \bibnamefont
  {Bohrdt}}, \bibinfo {author} {\bibfnamefont {Fabian}\ \bibnamefont {Grusdt}},
  \bibinfo {author} {\bibfnamefont {Eugene}\ \bibnamefont {Demler}}, \bibinfo
  {author} {\bibfnamefont {Martin}\ \bibnamefont {Lebrat}}, \ and\ \bibinfo
  {author} {\bibfnamefont {Markus}\ \bibnamefont {Greiner}},\ }\bibfield
  {title} {\enquote {\bibinfo {title} {Coupling a mobile hole to an
  antiferromagnetic spin background: Transient dynamics of a magnetic
  polaron},}\ }\href {\doibase 10.1103/PhysRevX.11.021022} {\bibfield
  {journal} {\bibinfo  {journal} {Phys. Rev. X}\ }\textbf {\bibinfo {volume}
  {11}},\ \bibinfo {pages} {021022} (\bibinfo {year} {2021})}\BibitemShut
  {NoStop}%
\bibitem [{\citenamefont {Koepsell}\ \emph {et~al.}(2021)\citenamefont
  {Koepsell}, \citenamefont {Bourgund}, \citenamefont {Sompet}, \citenamefont
  {Hirthe}, \citenamefont {Bohrdt}, \citenamefont {Wang}, \citenamefont
  {Grusdt}, \citenamefont {Demler}, \citenamefont {Salomon}, \citenamefont
  {Gross},\ and\ \citenamefont {Bloch}}]{Koepsell2021}%
  \BibitemOpen
  \bibfield  {author} {\bibinfo {author} {\bibfnamefont {Joannis}\ \bibnamefont
  {Koepsell}}, \bibinfo {author} {\bibfnamefont {Dominik}\ \bibnamefont
  {Bourgund}}, \bibinfo {author} {\bibfnamefont {Pimonpan}\ \bibnamefont
  {Sompet}}, \bibinfo {author} {\bibfnamefont {Sarah}\ \bibnamefont {Hirthe}},
  \bibinfo {author} {\bibfnamefont {Annabelle}\ \bibnamefont {Bohrdt}},
  \bibinfo {author} {\bibfnamefont {Yao}\ \bibnamefont {Wang}}, \bibinfo
  {author} {\bibfnamefont {Fabian}\ \bibnamefont {Grusdt}}, \bibinfo {author}
  {\bibfnamefont {Eugene}\ \bibnamefont {Demler}}, \bibinfo {author}
  {\bibfnamefont {Guillaume}\ \bibnamefont {Salomon}}, \bibinfo {author}
  {\bibfnamefont {Christian}\ \bibnamefont {Gross}}, \ and\ \bibinfo {author}
  {\bibfnamefont {Immanuel}\ \bibnamefont {Bloch}},\ }\bibfield  {title}
  {\enquote {\bibinfo {title} {Microscopic evolution of doped mott insulators
  from polaronic metal to fermi liquid},}\ }\href {\doibase
  10.1126/science.abe7165} {\bibfield  {journal} {\bibinfo  {journal}
  {Science}\ }\textbf {\bibinfo {volume} {374}},\ \bibinfo {pages} {82--86}
  (\bibinfo {year} {2021})},\ \Eprint
  {http://arxiv.org/abs/https://www.science.org/doi/pdf/10.1126/science.abe7165}
  {https://www.science.org/doi/pdf/10.1126/science.abe7165} \BibitemShut
  {NoStop}%
\bibitem [{\citenamefont {Gall}\ \emph {et~al.}(2021)\citenamefont {Gall},
  \citenamefont {Wurz}, \citenamefont {Samland}, \citenamefont {Chan},\ and\
  \citenamefont {K{\"o}hl}}]{gall2021a}%
  \BibitemOpen
  \bibfield  {author} {\bibinfo {author} {\bibfnamefont {Marcell}\ \bibnamefont
  {Gall}}, \bibinfo {author} {\bibfnamefont {Nicola}\ \bibnamefont {Wurz}},
  \bibinfo {author} {\bibfnamefont {Jens}\ \bibnamefont {Samland}}, \bibinfo
  {author} {\bibfnamefont {Chun~Fai}\ \bibnamefont {Chan}}, \ and\ \bibinfo
  {author} {\bibfnamefont {Michael}\ \bibnamefont {K{\"o}hl}},\ }\bibfield
  {title} {\enquote {\bibinfo {title} {Competing magnetic orders in a bilayer
  {{Hubbard}} model with ultracold atoms},}\ }\href {\doibase
  10.1038/s41586-020-03058-x} {\bibfield  {journal} {\bibinfo  {journal}
  {Nature}\ }\textbf {\bibinfo {volume} {589}},\ \bibinfo {pages} {40--43}
  (\bibinfo {year} {2021})}\BibitemShut {NoStop}%
\bibitem [{\citenamefont {Hirthe}\ \emph {et~al.}(2022)\citenamefont {Hirthe},
  \citenamefont {Chalopin}, \citenamefont {Bourgund}, \citenamefont
  {Bojovi{\'c}}, \citenamefont {Bohrdt}, \citenamefont {Demler}, \citenamefont
  {Grusdt}, \citenamefont {Bloch},\ and\ \citenamefont {Hilker}}]{Hirthe2022}%
  \BibitemOpen
  \bibfield  {author} {\bibinfo {author} {\bibfnamefont {Sarah}\ \bibnamefont
  {Hirthe}}, \bibinfo {author} {\bibfnamefont {Thomas}\ \bibnamefont
  {Chalopin}}, \bibinfo {author} {\bibfnamefont {Dominik}\ \bibnamefont
  {Bourgund}}, \bibinfo {author} {\bibfnamefont {Petar}\ \bibnamefont
  {Bojovi{\'c}}}, \bibinfo {author} {\bibfnamefont {Annabelle}\ \bibnamefont
  {Bohrdt}}, \bibinfo {author} {\bibfnamefont {Eugene}\ \bibnamefont {Demler}},
  \bibinfo {author} {\bibfnamefont {Fabian}\ \bibnamefont {Grusdt}}, \bibinfo
  {author} {\bibfnamefont {Immanuel}\ \bibnamefont {Bloch}}, \ and\ \bibinfo
  {author} {\bibfnamefont {Timon~A.}\ \bibnamefont {Hilker}},\ }\href {\doibase
  10.48550/ARXIV.2203.10027} {\enquote {\bibinfo {title} {Magnetically mediated
  hole pairing in fermionic ladders of ultracold atoms},}\ } (\bibinfo {year}
  {2022})\BibitemShut {NoStop}%
\bibitem [{\citenamefont {Chao}\ \emph {et~al.}(1977)\citenamefont {Chao},
  \citenamefont {Spalek},\ and\ \citenamefont {Oles}}]{Chao_1977}%
  \BibitemOpen
  \bibfield  {author} {\bibinfo {author} {\bibfnamefont {K~A}\ \bibnamefont
  {Chao}}, \bibinfo {author} {\bibfnamefont {J}~\bibnamefont {Spalek}}, \ and\
  \bibinfo {author} {\bibfnamefont {A~M}\ \bibnamefont {Oles}},\ }\bibfield
  {title} {\enquote {\bibinfo {title} {Kinetic exchange interaction in a narrow
  s-band},}\ }\href {\doibase 10.1088/0022-3719/10/10/002} {\bibfield
  {journal} {\bibinfo  {journal} {Journal of Physics C: Solid State Physics}\
  }\textbf {\bibinfo {volume} {10}},\ \bibinfo {pages} {L271--L276} (\bibinfo
  {year} {1977})}\BibitemShut {NoStop}%
\bibitem [{\citenamefont {Reischl}\ \emph {et~al.}(2004)\citenamefont
  {Reischl}, \citenamefont {{M{\"u}ller-Hartmann}},\ and\ \citenamefont
  {Uhrig}}]{reischl2004}%
  \BibitemOpen
  \bibfield  {author} {\bibinfo {author} {\bibfnamefont {Alexander}\
  \bibnamefont {Reischl}}, \bibinfo {author} {\bibfnamefont {Erwin}\
  \bibnamefont {{M{\"u}ller-Hartmann}}}, \ and\ \bibinfo {author}
  {\bibfnamefont {G{\"o}tz~S.}\ \bibnamefont {Uhrig}},\ }\bibfield  {title}
  {\enquote {\bibinfo {title} {Systematic mapping of the {{Hubbard}} model to
  the generalized t - {{J}} model},}\ }\href {\doibase
  10.1103/PhysRevB.70.245124} {\bibfield  {journal} {\bibinfo  {journal}
  {Physical Review B}\ }\textbf {\bibinfo {volume} {70}},\ \bibinfo {pages}
  {245124} (\bibinfo {year} {2004})}\BibitemShut {NoStop}%
\bibitem [{\citenamefont {Koepsell}\ \emph {et~al.}(2020)\citenamefont
  {Koepsell}, \citenamefont {Hirthe}, \citenamefont {Bourgund}, \citenamefont
  {Sompet}, \citenamefont {Vijayan}, \citenamefont {Salomon}, \citenamefont
  {Gross},\ and\ \citenamefont {Bloch}}]{koepsell2020}%
  \BibitemOpen
  \bibfield  {author} {\bibinfo {author} {\bibfnamefont {Joannis}\ \bibnamefont
  {Koepsell}}, \bibinfo {author} {\bibfnamefont {Sarah}\ \bibnamefont
  {Hirthe}}, \bibinfo {author} {\bibfnamefont {Dominik}\ \bibnamefont
  {Bourgund}}, \bibinfo {author} {\bibfnamefont {Pimonpan}\ \bibnamefont
  {Sompet}}, \bibinfo {author} {\bibfnamefont {Jayadev}\ \bibnamefont
  {Vijayan}}, \bibinfo {author} {\bibfnamefont {Guillaume}\ \bibnamefont
  {Salomon}}, \bibinfo {author} {\bibfnamefont {Christian}\ \bibnamefont
  {Gross}}, \ and\ \bibinfo {author} {\bibfnamefont {Immanuel}\ \bibnamefont
  {Bloch}},\ }\bibfield  {title} {\enquote {\bibinfo {title} {Robust {{Bilayer
  Charge Pumping}} for {{Spin-}} and {{Density-Resolved Quantum Gas
  Microscopy}}},}\ }\href {\doibase 10.1103/PhysRevLett.125.010403} {\bibfield
  {journal} {\bibinfo  {journal} {Physical Review Letters}\ }\textbf {\bibinfo
  {volume} {125}},\ \bibinfo {pages} {010403} (\bibinfo {year}
  {2020})}\BibitemShut {NoStop}%
\bibitem [{\citenamefont {Browaeys}\ and\ \citenamefont
  {Lahaye}(2020)}]{Browaeys:2020tl}%
  \BibitemOpen
  \bibfield  {author} {\bibinfo {author} {\bibfnamefont {Antoine}\ \bibnamefont
  {Browaeys}}\ and\ \bibinfo {author} {\bibfnamefont {Thierry}\ \bibnamefont
  {Lahaye}},\ }\bibfield  {title} {\enquote {\bibinfo {title} {Many-body
  physics with individually controlled rydberg atoms},}\ }\href {\doibase
  10.1038/s41567-019-0733-z} {\bibfield  {journal} {\bibinfo  {journal} {Nature
  Physics}\ }\textbf {\bibinfo {volume} {16}},\ \bibinfo {pages} {132--142}
  (\bibinfo {year} {2020})}\BibitemShut {NoStop}%
\bibitem [{\citenamefont {Reiter}(1994)}]{reiter1994c}%
  \BibitemOpen
  \bibfield  {author} {\bibinfo {author} {\bibfnamefont {George~F.}\
  \bibnamefont {Reiter}},\ }\bibfield  {title} {\enquote {\bibinfo {title}
  {Self-consistent wave function for magnetic polarons in the {\emph{t}} -
  {{{\emph{J}}}} model},}\ }\href {\doibase 10.1103/PhysRevB.49.1536}
  {\bibfield  {journal} {\bibinfo  {journal} {Physical Review B}\ }\textbf
  {\bibinfo {volume} {49}},\ \bibinfo {pages} {1536--1539} (\bibinfo {year}
  {1994})}\BibitemShut {NoStop}%
\bibitem [{\citenamefont {Ram{\v s}ak}\ and\ \citenamefont
  {Horsch}(1998)}]{ramsak1998c}%
  \BibitemOpen
  \bibfield  {author} {\bibinfo {author} {\bibfnamefont {A.}~\bibnamefont
  {Ram{\v s}ak}}\ and\ \bibinfo {author} {\bibfnamefont {P.}~\bibnamefont
  {Horsch}},\ }\bibfield  {title} {\enquote {\bibinfo {title} {Spatial
  structure of spin polarons in the t - {{J}} model},}\ }\href {\doibase
  10.1103/PhysRevB.57.4308} {\bibfield  {journal} {\bibinfo  {journal}
  {Physical Review B}\ }\textbf {\bibinfo {volume} {57}},\ \bibinfo {pages}
  {4308--4320} (\bibinfo {year} {1998})}\BibitemShut {NoStop}%
\bibitem [{\citenamefont {Ramsak}\ and\ \citenamefont
  {Horsch}(1993)}]{ramsak1993}%
  \BibitemOpen
  \bibfield  {author} {\bibinfo {author} {\bibfnamefont {A}~\bibnamefont
  {Ramsak}}\ and\ \bibinfo {author} {\bibfnamefont {P}~\bibnamefont {Horsch}},\
  }\bibfield  {title} {\enquote {\bibinfo {title} {Spin polarons in the t-{{J}}
  model: {{Shape}} and backflow},}\ }\href {\doibase 10.1103/PhysRevB.48.10559}
  {\bibfield  {journal} {\bibinfo  {journal} {Physical Review B}\ }\textbf
  {\bibinfo {volume} {48}},\ \bibinfo {pages} {4} (\bibinfo {year}
  {1993})}\BibitemShut {NoStop}%
\bibitem [{\citenamefont {Liu}\ and\ \citenamefont
  {Manousakis}(1992)}]{liu1992}%
  \BibitemOpen
  \bibfield  {author} {\bibinfo {author} {\bibfnamefont {Zhiping}\ \bibnamefont
  {Liu}}\ and\ \bibinfo {author} {\bibfnamefont {Efstratios}\ \bibnamefont
  {Manousakis}},\ }\bibfield  {title} {\enquote {\bibinfo {title} {Dynamical
  properties of a hole in a {{Heisenberg}} antiferromagnet},}\ }\href {\doibase
  10.1103/PhysRevB.45.2425} {\bibfield  {journal} {\bibinfo  {journal}
  {Physical Review B}\ }\textbf {\bibinfo {volume} {45}},\ \bibinfo {pages}
  {2425--2437} (\bibinfo {year} {1992})}\BibitemShut {NoStop}%
\bibitem [{\citenamefont {Vojta}\ and\ \citenamefont
  {Becker}(1999)}]{vojta1999}%
  \BibitemOpen
  \bibfield  {author} {\bibinfo {author} {\bibfnamefont {Matthias}\
  \bibnamefont {Vojta}}\ and\ \bibinfo {author} {\bibfnamefont {Klaus~W.}\
  \bibnamefont {Becker}},\ }\bibfield  {title} {\enquote {\bibinfo {title}
  {Doped bilayer antiferromagnets: {{Hole}} dynamics on both sides of a
  magnetic ordering transition},}\ }\href {\doibase 10.1103/PhysRevB.60.15201}
  {\bibfield  {journal} {\bibinfo  {journal} {Physical Review B}\ }\textbf
  {\bibinfo {volume} {60}},\ \bibinfo {pages} {15201--15213} (\bibinfo {year}
  {1999})}\BibitemShut {NoStop}%
\bibitem [{\citenamefont {Skou}\ \emph {et~al.}(2021)\citenamefont {Skou},
  \citenamefont {Skov}, \citenamefont {J{\o}rgensen}, \citenamefont {Nielsen},
  \citenamefont {{Camacho-Guardian}}, \citenamefont {Pohl}, \citenamefont
  {Bruun},\ and\ \citenamefont {Arlt}}]{skou2021}%
  \BibitemOpen
  \bibfield  {author} {\bibinfo {author} {\bibfnamefont {Magnus~G.}\
  \bibnamefont {Skou}}, \bibinfo {author} {\bibfnamefont {Thomas~G.}\
  \bibnamefont {Skov}}, \bibinfo {author} {\bibfnamefont {Nils~B.}\
  \bibnamefont {J{\o}rgensen}}, \bibinfo {author} {\bibfnamefont {Kristian~K.}\
  \bibnamefont {Nielsen}}, \bibinfo {author} {\bibfnamefont {Arturo}\
  \bibnamefont {{Camacho-Guardian}}}, \bibinfo {author} {\bibfnamefont
  {Thomas}\ \bibnamefont {Pohl}}, \bibinfo {author} {\bibfnamefont {Georg~M.}\
  \bibnamefont {Bruun}}, \ and\ \bibinfo {author} {\bibfnamefont {Jan~J.}\
  \bibnamefont {Arlt}},\ }\bibfield  {title} {\enquote {\bibinfo {title}
  {Non-equilibrium quantum dynamics and formation of the {{Bose}} polaron},}\
  }\href {\doibase 10.1038/s41567-021-01184-5} {\bibfield  {journal} {\bibinfo
  {journal} {Nature Physics}\ }\textbf {\bibinfo {volume} {17}},\ \bibinfo
  {pages} {731--735} (\bibinfo {year} {2021})}\BibitemShut {NoStop}%
\bibitem [{\citenamefont {Bruus}\ and\ \citenamefont
  {Flensberg}(2004)}]{bruus2004}%
  \BibitemOpen
  \bibfield  {author} {\bibinfo {author} {\bibfnamefont {Henrik}\ \bibnamefont
  {Bruus}}\ and\ \bibinfo {author} {\bibfnamefont {Karsten}\ \bibnamefont
  {Flensberg}},\ }\href@noop {} {\emph {\bibinfo {title} {Many-Body Quantum
  Theory in Condensed Matter Physics: An Introduction}}},\ Oxford Graduate
  Texts\ (\bibinfo  {publisher} {{Oxford University Press}},\ \bibinfo
  {address} {{Oxford ; New York}},\ \bibinfo {year} {2004})\BibitemShut
  {NoStop}%
\bibitem [{\citenamefont {Nielsen}\ \emph {et~al.}(2019)\citenamefont
  {Nielsen}, \citenamefont {Ardila}, \citenamefont {Bruun},\ and\ \citenamefont
  {Pohl}}]{Nielsen_2019}%
  \BibitemOpen
  \bibfield  {author} {\bibinfo {author} {\bibfnamefont {K~Knakkergaard}\
  \bibnamefont {Nielsen}}, \bibinfo {author} {\bibfnamefont {L~A~Pe{\~{n}}a}\
  \bibnamefont {Ardila}}, \bibinfo {author} {\bibfnamefont {G~M}\ \bibnamefont
  {Bruun}}, \ and\ \bibinfo {author} {\bibfnamefont {T}~\bibnamefont {Pohl}},\
  }\bibfield  {title} {\enquote {\bibinfo {title} {Critical slowdown of
  non-equilibrium polaron dynamics},}\ }\href {\doibase
  10.1088/1367-2630/ab0a81} {\bibfield  {journal} {\bibinfo  {journal} {New
  Journal of Physics}\ }\textbf {\bibinfo {volume} {21}},\ \bibinfo {pages}
  {043014} (\bibinfo {year} {2019})}\BibitemShut {NoStop}%
\end{thebibliography}%

\end{document}